\documentclass[twocolumn]{aastex631}

\newcommand{\flatiron}{\affiliation{Center for Computational Astrophysics, Flatiron Institute, New York, NY 10010, USA}}
\newcommand{\stonybrook}{\affiliation{Department of Physics and Astronomy, Stony Brook University, Stony Brook NY 11794, USA}}
\usepackage{amsmath}
\usepackage{amssymb}
\usepackage{bm}
\usepackage{xcolor}
\usepackage{showyourwork}
\definecolor{rb4}{HTML}{27408B}

\begin{document}
\title{Backward Population Synthesis: Mapping the Evolutionary History of Gravitational-Wave Progenitors}


\author{Kaze W. K. Wong}
\email{kwong@flatironinstitute.org}
\flatiron

\author{Katelyn Breivik}
\email{kbreivik@flatironinstitute.org}
\flatiron

\author{Will M. Farr}
\email{will.farr@stonybrook.edu}
\flatiron
\stonybrook

\author{Rodrigo Luger}
\email{rluger@flatironinstitute.org}
\flatiron

\date{\today}

\begin{abstract}
One promising way to extract information about stellar astrophysics from
gravitational wave catalogs is to compare the catalog to the outputs of stellar
population synthesis modeling with varying physical assumptions.  The parameter
space of physical assumptions in population synthesis is high-dimensional and
the choice of parameters that best represents the evolution of a binary system
may depend in an as-yet-to-be-determined way on the system's properties.  Here
we propose a pipeline to simultaneously infer zero-age main sequence properties
and population synthesis parameter settings controlling modeled binary evolution
from individual gravitational wave observations of merging compact binaries. Our
pipeline can efficiently explore the high-dimensional space of population
synthesis settings and progenitor system properties for each system in a catalog
of gravitational wave observations.  We apply our pipeline to observations in
the third third LIGO--Virgo Gravitational-Wave Transient Catalog. We showcase the
effectiveness of this pipeline with a detailed study of the progenitor
properties and population synthesis settings that produce mergers like the
observed GW150914.  Our pipeline permits a measurement of the variation of
population synthesis parameter settings with binary properties, if any; we
present inferences for the recent GWTC-3 transient catalog that suggest that the
stable mass transfer efficiency parameter may vary with primary black hole mass.  
\end{abstract}


\section{Introduction}
As the detection rate of gravitation waves (GWs) from merging
double-compact-object (DCO) binaries increases with the sensitivity of the
ground-based GW detector network
\citep{Aso2013,LIGOScientificCollaboration2015,Acernese2015,Abbott2018,Buikema2020,Tse2019,Acernese2019,Akutsu2021},
we are beginning to constrain the astrophysical processes which shape the
evolution of GW progenitor populations. One of the most common ways to study
progenitor populations of GW mergers is through population synthesis simulations
of stellar populations from a host of formation environments. In the case of isolated binary star evolution, merging
double compact objects can form from massive binary stars through the
standard stable mass transfer or common envelope channels
\citep[e.g.][]{Belczynski2002, Dominik2012, Belczynski2016, Stevenson2017, Zevin2020, Bavera2021, Broekgaarden2021, VanSon2022}
as well as through chemically homogenous evolution \citep{Mandel2016, Marchant2016, deMink2016}
or with population III stars \citep[e.g.][]{Belczynski2004, Kinugawa2014, Inayoshi2016, Inayoshi2017, Tanikawa2021, Tanikawa2022}.
Merging DCO binaries can also originate from a wide variety of dynamically active environments including
triple (or higher multiple) systems \citep[e.g.][]{Antonini2017, Silsbee2017, Fragione2019, VignaGomez2021},
as well as young stellar clusters \citep[e.g.][]{Ziosi2014, Banerjee2017, DiCarlo2020, Chattopadhyay2022},
globular clusters \citep[e.g.][]{PortegiesZwart2000, OLeary2006, Downing2010, Samsing2014, Rodriguez2015, Rodriguez2016, Askar2017, Rodriguez2019},
nuclear star clusters \citep[][]{Miller2009, Antonini2016}, or a mix of all three \citep[e.g.][]{Mapelli2022}.
Finally, more exotic channels like active galactic nuclei which combine both gravitational and gas interactions
\citep[e.g.][]{McKernan2018, McKernan2020, Secunda2020, Ford2021} or primordial black holes \citep[e.g.][]{Bird2016, AliHaimoud2017}
can produce merging DCOs observable with a ground-based detector network. For a discussion of the relative rates of each
formation environment and channels within, see \citet{Mandel2022} and references therein.

Traditionally, population synthesis studies simulate merging DCO populations
with a Monte-Carlo approach.  Using theoretically motivated distributions of
initial parameters like age, metallicity, component mass, orbital separation,
and eccentricity, initial populations are evolved with fixed astrophysical
assumptions, or population synthesis hyperparameter settings like the stability 
of Roche-overflow mass transfer, common envelope ejection efficiency, or the 
strength of compact-object natal kicks, to produce a
synthetic catalog of DCO mergers which can be compared to parameterized models
derived from observations. This approach relies on several assumptions, both in
the simulations and model parameterizations fitted to the GW data, which are
unlikely to be fully correct.  For example, there is no reason to suppose that
the entire population of DCOs should evolve with the same fixed hyperparameters;
there is no a priori reason the fitted parameterized model should capture the
relevant features of the observed DCO distributions to correspond to the binary
physics of interest; etc. A better approach would be to compare progenitor
parameters and hyperparameter settings directly to DCO observations, treating
the population synthesis model as a mapping from progenitor to merger
parameters; merger parameters can then be mapped into observations using a
gravitational waveform family, and parameter inference proceeds in the usual
way, propagating information back up the mapping chain
\citep[e.g.][]{Veitch2015}.  A similar methodology, though with \emph{fixed}
hyperparameter settings, was advanced in \citet{Andrews2018,Andrews2021}, as we
discuss in more detail below.

Attempting to approach such an improved inference procedure for evolutionary
hyperparameters using traditional en mass Monte Carlo simulations of entire
populations of DCO mergers is computationally infeasible. Even with recent
developments in using emulators to speed up the simulation process
\citep[e.g.][]{Wong2021}, training the emulators still requires a significant
amount of computation to cover a wide range of uncertainty in the hyperparemeter
space. As an example, to emulate a model with $10$ parameters, the simplest way
to construct a training set of simulations is to run simulations on a grid which
varies combinations of uncertain physics. For $10$ uncertain physical processes,
if we consider $2$ variations for each physical process this would still require
$1024$ simulations. Because population synthesis simulations require hundreds to
thousands of cpu hours, training an emulator which spans a high dimensional
uncertainty space remains an impossibility at present.  Thus inference of
hyperpameter settings must proceed \emph{dynamically}, generating trial DCO
systems via population synthesis recipes to match a particular observation one-
or few-at-a-time while simultaneously adjusting the hyperparameters. (Eventually
it may be even be possible to use more physically-motivated modeling of binary
physics \citep[e.g.][]{Gallego-Garcia2021} in such a procedure.)

\citet{Andrews2021} (hereafter A21) used \texttt{DartBoard} \citep{Andrews2018}
to determine the ZAMS parameters which produce GW150914-like BBH merger based on
posterior samples for GW150914 and a fixed set of hyperparameters which define
assumptions for how isolated binary-star interactions proceed using
\texttt{COSMIC}, a binary population synthesis code \citep{Breivik2020}. The
approach is very similar to the one advocated here \emph{except} that
\citet{Andrews2021} treated the evolutionary hyperparameters for the binary
physics as fixed within each analysis.  The key extension in this work, which
will enable population modeling of \emph{both} progenitor parameters and
hyperparemeter settings, including any possible \emph{dependence} of
hyperparameter settings on progenitor parameters, is to allow these
hyperparameters to vary at the same time as intrinsic properties of the
progenitor system to produce a joint inference over the intrinsic properties of
the progenitor and the necessary hyperparameter settings to produce the observed
merger properties.

In full: we propose a method to ``backward" model each GW event to its
progenitor state while allowing the hyperparameters to vary across their full
range of physical uncertainty following a two-stage process. First, we solve a
root finding problem to obtain guesses that are likely to produce the desired
system properties in the joint progenitor--hyperparameter space. We then sample
the posterior in the joint space that is induced by the DCO merger observations
using a Markov chain Monte Carlo (MCMC) algorithm that is initialized by the
roots found in the previous stage.  We demonstrate our algorithm in a realistic
setting by producing a posterior over progenitor properties and \texttt{COSMIC}
hyperparameters implied by the observation of GW150914.  

We find that the progenitor properties and even the ability to produce a
GW150914-like merger event is strongly correlated with hyperparameter settings;
different assumptions about the black hole masses at merger in GW150914 imply
wildly different formation channels and ZAMS progenitor masses for this event
and some nearby combinations of ZAMS masses are unable, with any reasonable
hyperparameter settings, to produce a DCO merger like GW150914.  We further
exhibit preliminary results of an ongoing analysis over the entire GWTC-3
catalog \citep{GWTC-3} that suggest that the accretion efficiency in stable mass
transfer may depend non-trivially on the primary black hole mass in merging
systems; our methodology allows for a systematic study of such dependencies,
which we will explore more fully in future work.

The rest of the paper is structured as
follows: We detail our method in Sec. ~\ref{sec:method}. In Sec.
~\ref{sec:result}, we show the results of applying our method to a number of
real events. We discuss the implications of this work and future direction in
Sec. ~\ref{sec:discussion}.

\section{Method}
\label{sec:method}

We define the progenitor's zero age main sequence (ZAMS) properties like
mass, orbital period, and eccentricity as progenitor parameters $\bm{\theta'}$, the parameters
that control different uncertain physical prescriptions such as wind strength and common
envelope efficiency as hyperparameters $\bm{\lambda}$, and random variables
that affect certain stochastic process such as whether CO natal kick unbinds a
binary as $\bm{X}$. To avoid clutter in the following derivation, we denote all
the parameters related to mapping a particular progenitor system into a GW event
collectively as evolutionary parameters $\bm{\Theta}$, which includes
$\bm{\theta'}$, $\bm{\lambda}$, and $\bm{X}$.

Once all the parameters including a particular draw of all the random variables
represented by $\bm{X}$ are known, the population synthesis code is a
deterministic function that transforms the properties of the progenitors into
the GW parameters $\bm{\theta}$:
\begin{equation}
    \bm{\theta} = \bm{f}\left( \bm{\Theta} \right).
\end{equation}
The mapping $\bm{f}$ from $\bm{\Theta}$ to $\bm{\theta}$ can be many-to-one
due to degeneracies in the different physical processes and initial parameters
of the GW event progenitors.
In order to draw inferences about $\bm{\Theta}$ from GW data, we
must be able to evaluate the likelihood of that data at fixed progenitor
parameters, namely
\begin{equation}
    p\left( d \mid \bm{\Theta} \right) = p\left( d \mid \bm{\theta} = \bm{f}\left( \bm{\Theta} \right) \right).
\end{equation}
The equality holds because the likelihood depends only on the gravitational
waveform generated by parameters $\bm{\theta}$ onto which the progenitor
parameters map.  In principle this likelihood could be computed at arbitrary
values of $\bm{\Theta}$ using the same machinery that is used to estimate source
parameters in GW catalogs \citep{Veitch2015,Ashton2019,Romero-Shaw2020,GWTC-3}
to develop samples of progenitor parameters, hyperparameters, and random
variables in $\bm{\Theta}$ from a posterior density.

Since the GW likelihood function
ultimately depends only on the GW parameters $\theta$, and GWTC-3 already has samples over these parameters from a posterior density
\begin{equation}
    p\left( \bm{\theta} \mid d \right) \propto p\left( d \mid \bm{\theta} \right) \pi_\mathrm{GW} \left( \bm{\theta} \right),
\end{equation}
where $\pi_\mathrm{GW}$ is the prior density used for sampling, we can save the
computational cost associated with evaluating the GW likelihood by rewriting the
posterior density of $\bm{\Theta}$ in terms of the posterior density of
$\bm{\theta}$, namely

\begin{align}
    p(\bm{\Theta} | d) = \frac{p(\bm{\theta}(\bm{\Theta})| d) \pi(\bm{\Theta})}{\pi_\mathrm{GW}(\bm{\theta}(\bm{\Theta}))}.
\end{align}
We apply a kernel density estimator with a Gaussian kernel to the posterior
samples released in GWTC-3 to estimate $p(\bm{\theta}(\bm{\Theta})|d)$.

Including hyperparameters and random variables 
drastically increases the dimensionality of the problem, which can make the
sampling process much more computationally expensive to converge. To speed up
the convergence, we first solve a root-finding problem to find points
$\bm{\Theta}$ that will generate GW parameters $\bm{\theta}$ close to the bulk
of the posterior density of a GW event.  We then use those points to initialize
a set of chains in the MCMC process.

For each posterior sample point in the GW-observable space,
we can find the corresponding evolutionary parameters by solving an optimization problem that minimizes
the mean square difference between the GW event parameters and evolutionary parameters as:

\begin{align}
\mathcal{L}(\bm{\Theta},\bm{\theta}) = ||f(\bm{\Theta})-\bm{\theta}||^2.
\label{eq:loss}
\end{align}

In principle, we should only accept solutions that exactly reproduce the LVK posterior samples in the GW-observable space.
However, it is not feasible to achieve such a condition in practice, therefore we relax the condition in eq.~\ref{eq:loss}
to a small acceptance threshold. For this study, we picked a threshold of $10^{-2}$. To make sure we find a reasonably
complete set of progenitor parameters that corresponds to the posterior sample point, we use 1000 different initial guesses
in solving the optimization problem. As long as the solution fulfills the acceptance criteria, the root is as valid as all
other roots that fulfill the same acceptance criteria, \emph{regardless to its initial guess}.
This means, we have the freedom to choose the set of initial guesses however we think might benefit the optimization.

Most of the points in the evolutionary parameter space do not produce DCO mergers in a Hubble time. For systems that do not merge,
we set the binary masses to be $0$ such that the gradient for the same systems is also $0$ and leads the root finder to get stuck on
its first step.{\footnote{On a plateau of constant loss, the root finder performs no better than a random guess.
Because of this, a root finder in a high dimensional space will require a significant amount of computation
to get close to a solution.}} Therefore, it is beneficial to choose initial guesses in region of the evolutionary parameter
space that are likely to produce DCO mergers. To construct a list of initial guesses, we evolve a set of
ZAMS parameters uniformly sampled from the initial binary parameter space, then keep the systems that
merge within a Hubble time.

Retaining explicit control over random variables allows us to marginalize over
their contribution, and thus allows us to focus on progenitor parameters and
hyperparameters. The strength and direction of natal kicks for compact objects 
are directly specified in the \texttt{COMPAS} binary population synthesis code 
\citep{Riley2022} and can also be specified in \texttt{COSMIC}, however 
this requires the user to specify the strength and direction of natal kicks as 
inputs through a user-specified pre-processing script.  In this study, we use 
\texttt{COSMIC} in it's default state which does not specify the kick strength 
and direction at run time. 
This means that the process of evolving the binary
is not fully deterministic. Thus, even if the root-finding algorithm performs
perfectly, forward modelling a set of roots does not guarantee that the
simulated population reproduces exactly the set of posterior samples due to
randomness in the evolution of each binary. To assure that the recovered
progenitor and hyperparameters robustly correspond to the posterior in the
GW-observable space, we push forward (or ``reproject'') the recovered
evolutionary parameters to the GW-observable space to check whether the
reprojected posterior agrees with the posterior given by the LVK collaboration.


We use KL divergence to measure the agreement between the two posterior
distributions as:
\begin{align}
D_{KL}(P||Q) = \int P(\bm{x}) \log(P(\bm{x})/Q(\bm{x})) dx.
\label{eq:KLdivergence}
\end{align}

\noindent A small KL divergence means the reprojected \texttt{COSMIC} posterior is similar to the original posterior in the
observable space, and is thus a viable channel for that specific event. Otherwise, it either means \texttt{COSMIC}
can only explain part of the posterior or cannot explain the event at all. This is completely expected behavior since
\texttt{COSMIC}, and isolated binary evolution more generally, carries its own assumptions and is expected to
fail in reproducing a subset of the event in GWTC-3. One example of this is GW events with at least one component
with mass above the pair instability supernova mass limit \citep{Woosley2017, Farmer2019}.
The reprojection could also be subject to stochasticity in the evolution of each binary.
To account for this, we reproject the posterior multiple times with different random seeds and
check whether the KL divergence varies significantly. A varying KL divergence means a particular GW event is subject
to randomness in the evolution of the progenitor, and extra caution should be used when interpreting the result.

\section{Results}
\label{sec:result}

\begin{table*}[hbt!]
    \begin{center}
    \begin{tabular}{ l l l }
    \hline
    \hline
    Parameters &  Description & Optimization range\\
    \hline
    \hline
    Observables &\ &\  \\
    \hline
    \hline
    $m_{\rm 1, GW}$ & Primary mass of the GW event & NA \\
    $m_{\rm 2, GW}$ & Secondary mass of the GW event  & NA\\
    \hline
    \hline
    Progenitor parameters &\ &\  \\
    \hline
    \hline
    $m_{\rm 1, ZAMS}$ & Primary mass at ZAMS & $[10,150]\ M_{\odot}$\\
    $m_{\rm 2, ZAMS}$ & Secondary mass at ZAMS & $[10,150]\ M_{\odot}$\\
    $t_{\rm orb}$ & Orbital period at ZAMS & $[5,5000]$ days\\
    $e$ & Eccentricity at ZAMS & $[0,1]$\\
    $Z$ & Metallicity at ZAMS & $[10^{-4},2\times10^{-2}]$\\
    $z$ & Redshift at ZAMS & NA\\
    \hline
    \hline
    hyperparameters &\ &\ \\
    \hline
    \hline

    $\alpha$ & Common envelope efficiency & $[0.1,10]$\\
    $f_{\rm acc}$ & Fraction of accretion during stable mass transfer & $[0,1]$\\
    $q_{\rm crit,3}$ & Critical mass ratio on the Hertzsprung Gap & $[0.1,10]$\\
    $\sigma$ & Root mean square of Maxwellian natal kick distribution& $[10,300]\ {\rm km/s}$\\

    \hline
    \hline
    \end{tabular}
    \caption{A list of parameters used in this study.}
    \label{tab:parameters}
    \end{center}
\end{table*}

We apply the method described in section \ref{sec:method} to the first GW event GW150914 \citep{GW150914}.
We use the \texttt{Overall\_posterior} posterior samples for GW150914,
publicly available from the Gravitational wave Open Science Center \citep{LIGOScientific:2019lzm}. 
The parameters involved in the analysis and the range we allow for the inference are tabulated in table \ref{tab:parameters}.
The spin of BHs formed in isolated binaries depends strongly on the angular
momentum transport in massive stars which is still highly uncertain (see however \citet{Fuller2019, Bavera2020}),
therefore we only consider the two component masses in the observable space in both the
root-finding stage and the MCMC stage. For progenitor parameters, we characterize each guess with
five parameters: the two component masses $M_{\rm 1,ZAMS}$ and $M_{\rm 2,ZAMS}$, the orbital period $t_{\rm orb}$,
the eccentricity $e$, and the metallicity $Z$ at ZAMS. Note that the progenitor formation redshift is not
obtained through the root-finding process because the formation redshift does not intrinsically affect
the evolution of the binary. Instead, we compute the time it takes for the binary to evolve from
formation to the merger and add this time to the lookback time when a GW event is observed and limit the total
time to merger to be less than $13.7\,\rm{Gyr}$.
Using the total lookback time, we can compute the redshift of the binary at ZAMS formation following the
Planck 2018 cosmological model \citep{Planck2018}.
The main benefit of post-processing the redshift in this way is that we do not rely on a particular assumption
of star formation rate (SFR) distribution when we solve for the posterior distribution in redshift.
This means we can put the prior on formation redshift (or in general, metallicity-redshfit distribution) in
by reweighting the posterior distribution according to a particular SFR model,
and therefore do not have to rerun the inference every time we change the SFR model. This is especially helpful
since SFR models are still highly uncertain but strongly impact the local merger rates of BBHs \citep[e.g. ][]{Broekgaarden2021}.
For hyperparameters, we choose parameters that strongly affect
binary evolution in \texttt{COSMIC} for massive stars, including the common envelope
efficiency $\alpha$, the fraction of mass accreted onto the accretor during stable mass transfer $f_{\rm acc}$,
the critical mass ratio which determines whether Roche-overflow mass transfer
proceeds stably or produces a common envelope when the donor star is on the Hertzsprung gap
$q_{\rm crit,3}$, and the root-mean-square of the Maxwellian distribution
employed for CO natal kicks, $\sigma$. For systems which begin a common envelope 
with a Hertzsprung-gap donor, we always assume the pessimistic outcome that the two 
stars merge since the donor has likely not formed a strong core-envelope boundary 
\citep{Ivanova2004, Belczynski2008}.

\begin{figure*}[h]
    \includegraphics[width=\textwidth]{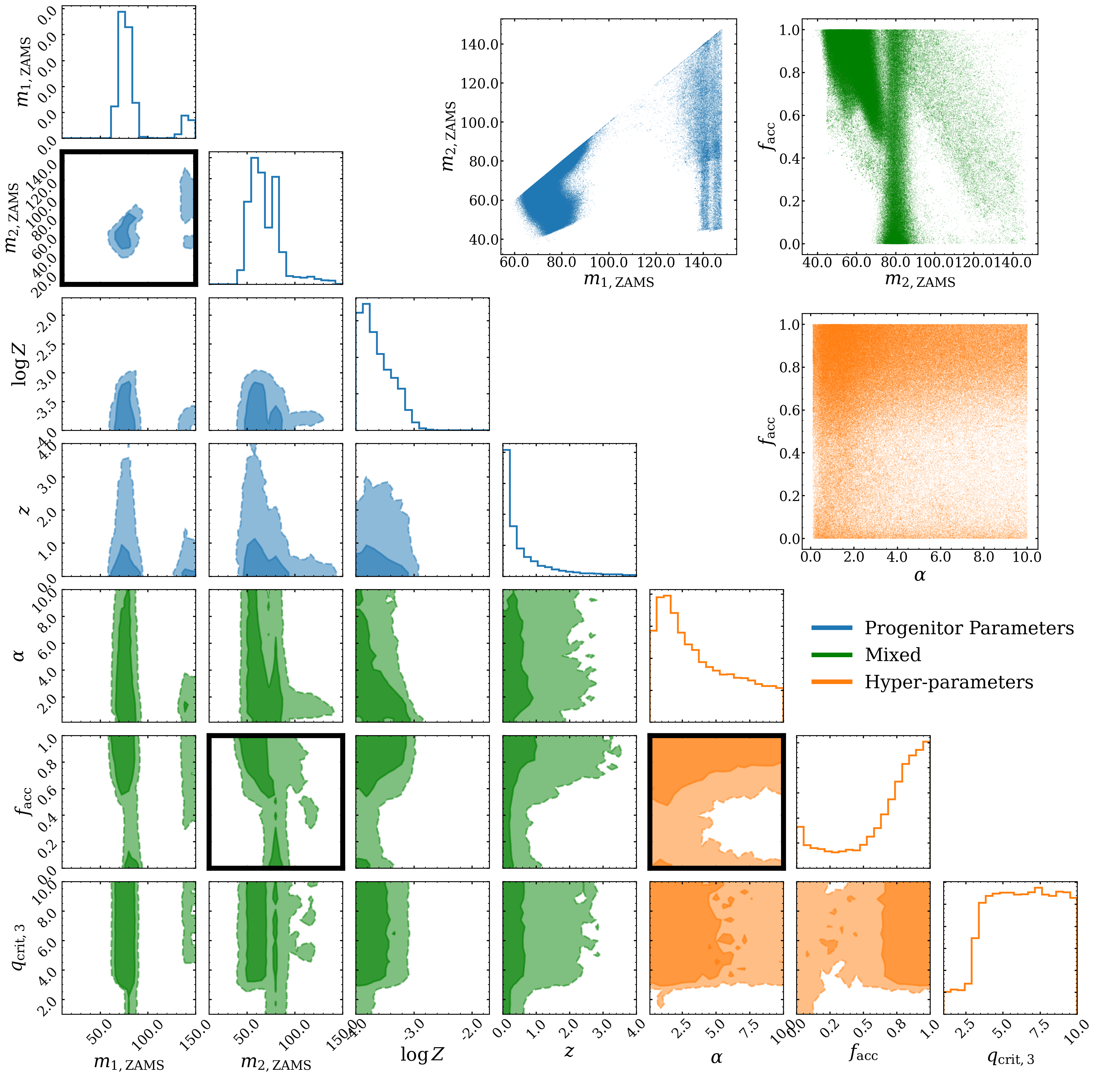}
    \caption{The posterior for GW150914 in both the progenitor parameter space and hyperparameter space.
    $M_1$ and $M_2$ are the progenitors' masses. $\log{Z}$ is log metallicity at ZAMS.
    $z$ is the redshift at ZAMS.
    $\alpha$ is the common envelope efficiency.
    $f_{\rm acc}$ is the fraction of mass accreted during stable mass transfer.
    $q_{\rm crit, 3}$ is the critical mass ratio on the Hertzsprung Gap.
    The contours correspond to the $68\%$ and $95\%$ confidence intervals.
    Note that the redshift is not fitted during the root finding process or the MCMC process.
    Once we find the evolutionary parameters, we add the delay time to the lookback time of the observed posterior sample,
    then from the total lookback time we can compute the redshift at ZAMS.
    We highlight three panels in the corner plots to show the fine structure of the set of posterior samples in the evolutionary parameter space.
    We also color the posterior in a particular panel according to the type of parameters involved in the corner plot.
    Blue denotes panels that include only progenitor parameters,
    green denotes panels that include a mix of progenitor parameters and hyperparameters,
    and orange denotes panels that include only hyperparameters.
    }
    \label{fig:GW150914_posterior}
    \script{figure1_cornerPlot.py}
\end{figure*}

We show a portion of the joint posterior of the progenitor parameters and
hyperparameters of GW150914 in figure \ref{fig:GW150914_posterior}. Each panel
is colored based on the comparison between progenitor parameters (blue),
hyperparameters (orange), or a mix of the two (green). We find that the choice
for $\sigma$ does not affect the formation of GW150914-like BBHs. This is not
unexpected since we use the \citet{Fryer2012} delayed model which reduces the
natal kick strength at BH formation based on the amount of fallback onto the
proto-CO. In the case of GW150914-like BBHs, the BH progenitors are massive
enough that the fallback reduces the natal kick to zero. Because of this we do
not include $\sigma$ in Figure~\ref{fig:GW150914_posterior}. Similarly, we do
not include the ZAMS orbital period and eccentricity, which are correlated with
one another, but not strongly with the other progenitor parameters or
hyperparameters. We find strong correlations between the ZAMS masses, as well as
strong correlations between $M_{2,\rm{ZAMS}}$ and $f_{\rm{acc}}$, and
$f_{\rm{acc}}$ and $\alpha$. We show scatter plots of each of these combinations
in the upper-right inset of the figure.

The correlations between the ZAMS masses are similar to those found by A21 with
the majority of the population preferring primary and secondary masses between
$60$--$90\,M_{\odot}$. However, we also find some ZAMS masses which extend up to
$150\,M_{\odot}$ which is the limit imposed by our assumptions. We explore the
formation scenarios which lead to successful GW150914-like mergers and those
which fail to produce GW150914-like mergers below in
Figure~\ref{fig:compare_fixed_variable}.

The correlations between $M_{2,\rm{ZAMS}}$ and $f_{\rm acc}$ illustrate the
variety of ways which GW150914-like mergers can be produced. For binaries with
$M_{2,\rm{ZAMS}} < 70\,M_{\odot}$, we find that accretion efficiencies of
$f_{\rm acc} > 0.5$ are preferred. One reason for this is based on the
requirement that the total mass in the binary must remain above the total mass
of GW150914 and strongly non-conservative mass transfer (i.e. $f_{\rm acc} <
0.5$) reduces the total mass of the system. A compounding factor is that as mass
leaves the binary due to non-conservative mass transfer, the evolution of the
binary's orbit is less dramatic and leads to wider binaries on average and thus
fewer mergers in a Hubble time. For binaries with $M_{2,\rm{ZAMS}} \sim
80\,M_{\odot}$, $f_{\rm acc}$ is less constrained. This is because of a
preference for these secondaries to also have primary masses near $
80\,M_{\odot}$ and thus enter a double-core common envelope evolution in which
both stars' envelopes are ejected, leaving behind two stripped helium cores in a
tight orbit. In this case, $f_{\rm acc}$ does not affect the binary evolution
and is thus unconstrained. This effect can also be seen in the ZAMS mass and
$q_{\rm crit, 3}$ panels for masses near $80\,M_{\odot}$ because double-core
common-envelope evolution is triggered in \texttt{COSMIC} when the radii of the
two stars touch due to the rapid expansion of the primary star upon helium
ignition. In this case, the choice for Roche-overflow mass transfer to proceed
stably (as prescribed by $f_{\rm{acc}}$) or unstably (as prescribed by $\alpha$)
using $q_{\rm crit, 3}$ is totally irrelevant, and thus unconstrained. Finally,
for binaries with $M_{2,\rm{ZAMS}} > 80$, we find that $f_{\rm acc}$ is
correlated to decrease with increasing mass. This is due to limits on the total
mass which must be ejected from the binary to produce BH masses that match
GW150914, the reverse situation to $M_{2,\rm{ZAMS}} < 70 \, M_\odot$. Binaries
with both component masses near $150\,M_{\odot}$ must either go through a common
envelope evolution (in which case $f_{\rm acc}$ is unconstrained), or very
non-conservative mass transfer (where $f_{\rm acc}$ is low), to produce BBHs
with the proper mass.

Finally, we find that $\alpha$ and $f_{\rm acc}$ are largely uncorrelated,
though there exist independent trends in each hyper parameter. This is not
totally unexpected since the physical processes which are described by each
hyperparameter, i.e. stable Roche-lobe overflow and common envelope evolution
are two independent channels. Generally, we find that GW150914-like BBHs tend to
prefer larger accretion efficiencies and have common envelope ejection
efficiencies peaking near $\alpha=1$. One shortcoming of our method
implementation is the application of a single hyper parameter for
$f_{\mathrm{acc}}$ and $\alpha$ for each binary, while both the primary and the
secondary star could in principle be defined by their own hyperparameters that
prescribe the outcomes for when each star fills it's Roche lobe. We reserve full
treatment of this for future work but note that this improvement could reveal
correlations between hyperparameters that are not present in our current
analysis.

Once we obtain a set of binaries which successfully map the ZAMS parameters and
hyperparameters to BBH merger masses, we re-evolve the set of ZAMS parameters
with the same physical assumptions as A21, but vary the common envelope
efficiency to explore how keeping a fixed model which only varies one
hyperparameter contrasts to our results.
Figure~\ref{fig:compare_fixed_variable} shows the distribution of ZAMS masses
for three models (first through third columns), each with a different $\alpha$
but the same hyperparameters as A21, as well as the results of our sampling
which allows our hyperparameters to vary (fourth column). We find that holding
the accretion efficiency, $f_{\rm{acc}}$, and common envelope ejection
efficiency, $\alpha$, to fixed values greatly reduces the ZAMS parameter space
that produces GW150914-like mergers. In contrast, by allowing the
hyperparameters to fully span the model uncertainty, we find that there are
distinct ZAMS parameters which produce GW150914-like mergers.

Because we can explore the full evolutionary parameter space, we can also
determine that there are ZAMS parameters which fail to produce GW150914-like
mergers \emph{regardless of our hyperparameter choice.} For primary ZAMS masses
between $\sim95-135\,M_{\odot}$ and secondary ZAMS masses between
$\sim50-95\,M_{\odot}$, we find that there is no combination of accretion
efficiency and common envelope ejection efficiency which produces merging BBHs
that have masses consistent with GW150914's posteriors. In this region, if BBHs
with masses consistent with GW150914's masses are produced through a combination
of $\alpha$ and $f_{\rm{acc}}$, they do not merge in a Hubble time. Conversely,
in the region of the fourth column of Figure~\ref{fig:compare_fixed_variable}
which does produce GW150914-like mergers, the combination of $\alpha$,
$f_{\rm{acc}}$, and the ZAMS masses and orbital separations balance to produce
BHs with the correct masses that merge within a Hubble time.

\begin{figure*}
    \includegraphics[width=0.98\textwidth]{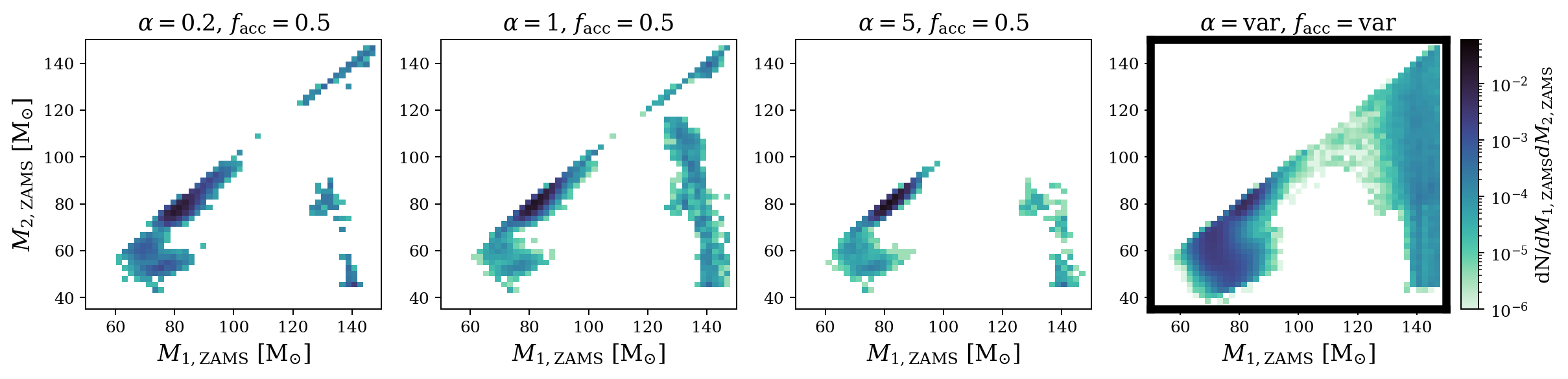}
    \caption{Comparison of ZAMS masses for binaries which produce GW150914-like mergers for three variations of $\alpha$
     with a fixed set of parameter assumptions matching those of A21 (first three columns) and for binaries which produce
     GW150914-like mergers when $\alpha$, $f_{\mathrm{acc}}$, and $q_{crit,3}$ are allowed to vary, (fourth column).}
    \label{fig:compare_fixed_variable}
    \script{figure2_varyAlphaFacc.py}
\end{figure*}

\begin{figure}
\includegraphics[width=0.49\textwidth]{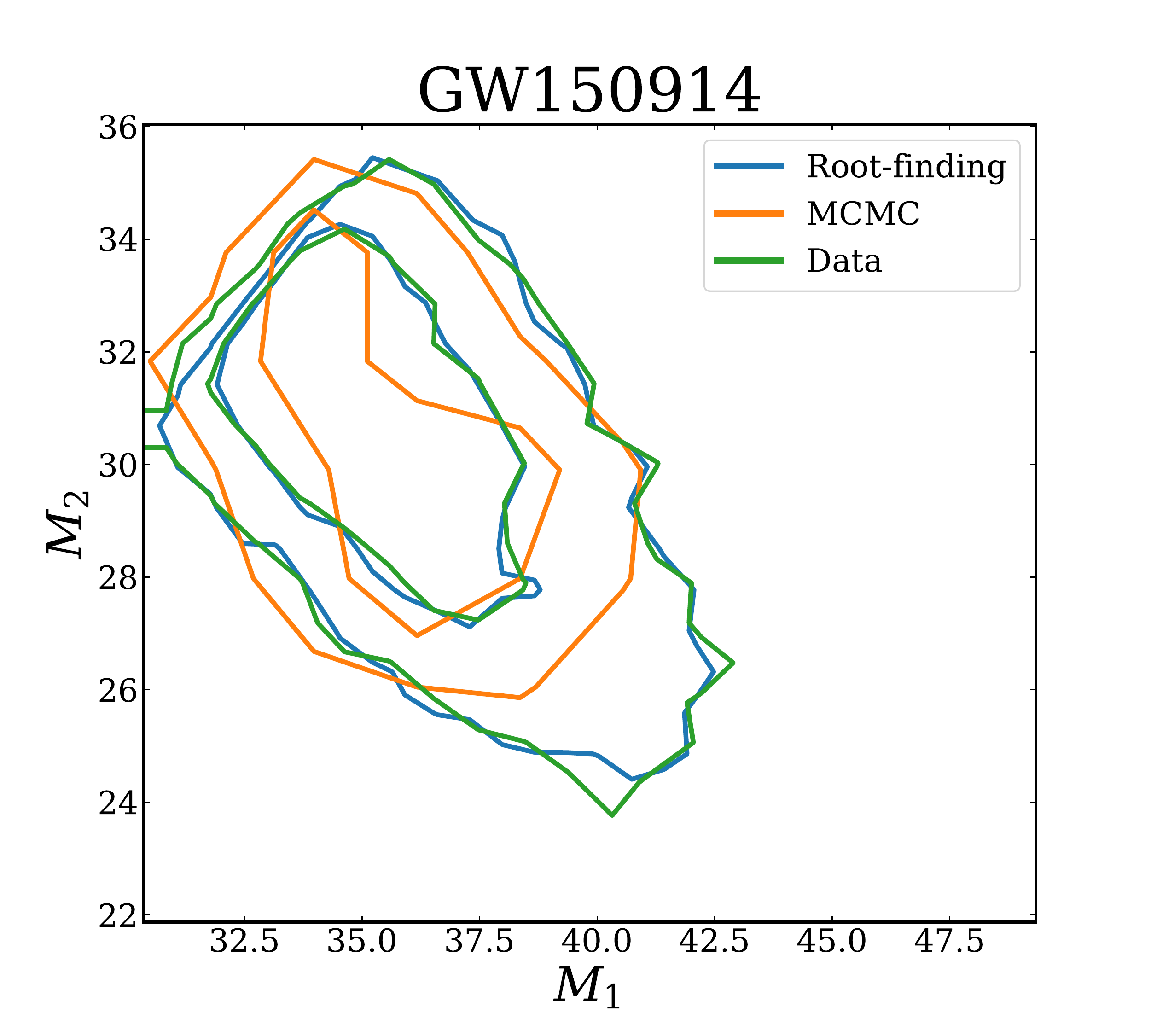}
\caption{Reprojecting the posterior in evolutionary parameter space of GW150914 to observable space.
The contours correspond to the $68\%$ and $95\%$ confidence intervals.
The blue contour is the reprojected posterior after the root-finding procedure.
The orange contour is the reprojected posterior after the MCMC procedure.
The green contour is the posterior plotted using the original LVK posterior samples.
}
\label{fig:GW150914_reprojection}
\script{figure3_sanityCheck.py}
\end{figure}

It is interesting to re-project the posterior over progenitor parameters ($\bm{\theta}'$)
and hyperparameters ($\bm{\lambda}$) using fresh random variables ($\bm{X}$) to
the observable space of GW parameters ($\bm{\theta}$).  Agreement between the
observed parameters and the re-projected parameters indicates good exploration
of the random variable space and lack of sensitivity to the details of these
random parameters.  Figure \ref{fig:GW150914_reprojection} shows good agreement
between the re-projected root-finding outputs as well as the re-projected MCMC
draws for our analysis of GW150914.  Unlike \citet{Andrews2021}, we find there
are no regions of observable space inaccessible to our evolutionary models; this
difference arises because we allow the evolutionary hyperparameters
$\bm{\lambda}$ to vary.

\begin{figure}
\includegraphics[width=0.49\textwidth]{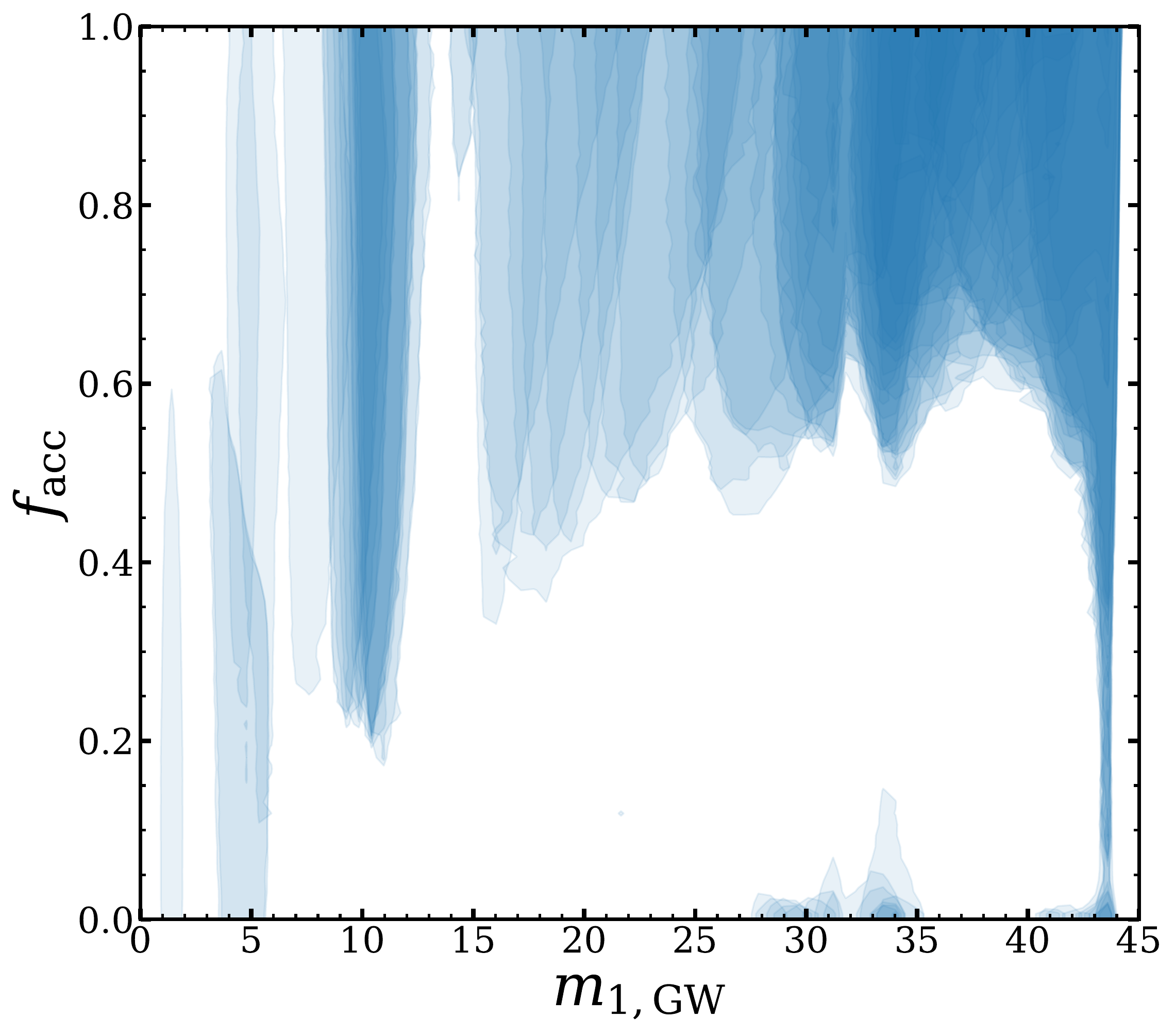}
\caption{
    The posterior density in the $m_{1,{\rm GW}}-f_{\rm acc}$ space for most of the events in GWTC-3.
    Note that $m_{1,{\rm GW}}$ here refers to the reprojected posterior instead of the posterior released by LIGO.
    Each contour is the $68\%$ credible interval of the posterior density for a particular event.
    At $m_{1,{\rm GW}} \sim 45 M_{\odot}$, the pair instability supernova mechanism prevents \texttt{COSMIC} from producing events that are more massive than this cutoff.
    Therefore, events with the majority of posterior support above this cutoff are not compatible with \texttt{COSMIC}, hence have a large KL divergence and are excluded from this figure.
    On the low mass end, neutron star binaries or neutron star-black hole binaries are subjected to randomness induced by the natal kick,
    also resulting in a larger and fluctuating KL divergences, and therefore are also excluded from the analysis.
}
\label{fig:GWTC-3_f_acc_mass}
\script{figure4_population.py}
\end{figure}

To illustrate the potential benefit of using our method on the population level,
we perform the same analysis for all events in GWTC-3.
For events announced in GWTC-1, we use \texttt{Overall\_posterior}, \texttt{PublicationSamples},\texttt{IMRPhenomXPHM\_comoving} and \texttt{C01:Mixed} posterior samples, respectively.
All the data are publicly available from the Gravitational wave Open Science Center \citep{LIGOScientific:2019lzm}.
Plotted in figure
\ref{fig:GWTC-3_f_acc_mass} is the posterior density in the $m_{1,{\rm
GW}}-f_{\rm acc}$ space for most of the events in GWTC-3.
Each contour represents
the $68\%$ credible interval of the posterior density for that particular event.
There is a suggestive trend showing that $f_{\rm acc}$ could increase as the mass of
the progenitor increases. Such a trend implies the stable mass transfer
phase of a less massive binary would be preferentially non-conservative, with
more conservative mass transfer in more massive binary systems.

Note that some events did not pass the KL divergence test we proposed in section
\ref{sec:method}; we do not include these systems in figure
\ref{fig:GWTC-3_f_acc_mass}. Binaries that form low-mass compact objects have
lower amounts of fallback and tend to have correspondingly larger variance in
their random natal kicks. Larger kicks can unbind the progenitor binary, or lead
to wide binaries that do not merge within a Hubble time.  In these cases, the
extra variance during the reprojection can produce a posterior that may not
agree with the original posterior, hence yielding a higher KL divergence.

For high-mass binaries, \texttt{COSMIC} struggles to produce events above the
pair instability supernova mass cutoff, so the posterior in the
evolutionary parameter space only corresponds to part of the posterior in the
observable space below this cutoff, and therefore events beyond the lower edge of the PISN mass
gap also have higher KL divergence.

Events with more extreme mass ratios are hard to produce with \texttt{COSMIC},
and therefore less likely to be accurately recovered by our method \citep[e.g.][]{Zevin2020}. This could
be due to our method assuming a single value for $f_{\rm acc}$ and $\alpha$ as
discussed in Section~\ref{sec:result}.

Merger observations for which \texttt{COSMIC} struggles are likely to be very
informative about formation channels and binary evolution physics, and are
therefore likely worthy of close study.  We anticipate that future work will
follow up these events in detail.

Figure \ref{fig:GWTC-3_f_acc_mass} shows that our method could in principle
reveal the correlation between progenitor parameters and hyperparameters on a
population level. Here we impose a cut to eliminate events with KL divergence larger
than 0.1.
This is a heuristic choice to exclude events that are obviously not compatible with
\texttt{COSMIC}.
A careful treatment of all events in the catalog and
discussion related to the detailed physical implications of figure
\ref{fig:GWTC-3_f_acc_mass} is beyond the scope of this paper.  We defer a
detailed study of the physics related to the population of GW events to future
work.

\section{Discussion}
\label{sec:discussion}

We present a new pathway to understand binary evolution with GW events in this
paper. Instead of forward modelling an assumed distribution of initial binaries
to the observed population, we find the corresponding evolutionary parameters
event-by-event. In this first work, we showcase the power of the proposed method
with an application to GW150914. We show the joint posterior of the event's
progenitor parameters and hyperparameters.  Our work is both a more efficient
way to study GW event progenitors, and also allows the possibility of
constraining the astrophysics related to binary evolution, especially by
capturing the correlation between hyperparameters in different systems.

Our results suggest that the accretion efficiency during stable mass transfer
may depend on the primary black hole mass.  In general, our method returns the
joint posterior distribution of progenitor parameters and hyperparameters for
each event, which enables a data-driven way to study the distribution of
hyperparameters.  That is, once we have a catalog of events, each has their own
posterior in the hyperparameter space, and we can employ well-known techniques
such as hierarchical Bayesian analysis to fit a population model to the
distribution of hyperparameters; this avoids making overly-specific assumptions
like fixing the hyperparameters for all types of event.

Working event-by-event as we do permits post-processing application of
arbitrarily complicated models of star formation and metalliticy evolution by
re-weighting the samples we obtain \citep[e.g.][]{VanSon2022}. Once we have
pulled the GW event posterior back from the observable space to the evolutionary
parameter space, we can apply the same hierarchical inference methods currently
used to infer the compact binary population to our progenitor population.  In
particular, we can use our progenitor population to infer the formation rate of
compact binary population progenitor systems over cosmic time without needing to
make artifical and simplistic assumptions about the delay time distribution or
the metallicity-specific star formation rate
\citep{Vitale2019,Ng2021,VanSon2022}.  Comparing the formation rate of
progenitors to the star formation rate provides an avenue to check our
understanding of binary evolution and the relative rate of other formation
channels.

While our method allows data-driven exploration of the hyperparameters space for the first time, there are a number of improvements that can be implemented in future studies.
In this study we use only \texttt{COSMIC} as our evolutionary function, which by design cannot explain all the events in GWTC-3.
For example, events with either of the component mass larger than the lower edge of PISN gap such as GW190521 cannot be explained by \texttt{COSMIC}.
Alternative channels such as dynamical formation might be needed to explain some subset of the events in GWTC-3 \citep{Zevin2021}.
As the sensitivity of GW detectors increases, we expect to see more and more events that are unusual in some way.
Therefore, having a self-consistence population synthesis code that contains multiple formation channels is essential to accommodate the growing catalog of GW events.
In this paper, our main focus is to illustrate the concept of "back-propagating" GW-event posterior samples,
highlighting the capability of our method and motivating the benefits of building population synthesis codes that can work seamlessly with our method.
To avoid cluttering of focus, we discuss the physical implications of the results presented in this work under our specific assumptions (i.e. using \texttt{COSMIC} as our evolutionary function) in a future companion paper.

Due to the implicit definition of random variables in \texttt{COSMIC}, our evolutionary function is stochastic.
This introduces significant inefficiency in our root-finding and sampling algorithm.
The main stochasticity in \texttt{COSMIC} comes from natal kick, which significantly affect the evolutioary pathway of low mass events such as BNS mergers.
The effect of the natal kick is suppressed for heavier mass events due to fallback.
This means events with lighter masses are subject to stochasticity of the function,
where the sampling process for heavier events behaves as if the evolutionary function we use is deterministic.
Due to computational limitation, we only try 1000 different initial guesses per posterior sample in the root-finding process.
This means any posterior sample that has a probability of merging rarer than 1 in 1000 could be missed.
Obviously the problems which comes with the randomness can be alleviated by performing more tries per posterior sample,
but this is not scalable in practice.
On top of limitation in efficiency, some formation channels require explicit control of random variables by construction.
For example, in a dynamical formation scenario such as binaries that form in a globular cluster,
each binary has some probability of undergoing a multi-body encounter with another member in the cluster.
These encounter probability distributions are either studied with direct N-body simulations or semi-analytical methods.
In both cases, each member of the cluster is no longer completely independent of the other members, but coupled through the encounter probability distribution.
By studying the encounter probability distribution, we can infer the properties of the environment which the binary lives in.
This can only be done if we have explicit control over the random variables that characterize the encounter probability distribution.

Another technical note is that we use finite differencing to estimate the gradient of the objective function, which could be a significant source of error near transition points in the evolutionary parameter space.
Also, finite differencing is increasingly inefficient as we increase the dimensionality of the problem.
To improve the accuracy and efficiency in estimating the gradient of the objective function, automatic differentiation is a promising feature that modelers should consider incorporating in their population synthesis codes in the future.

To summarize, we propose a novel method to recover the posterior samples in the evolutionary parameter space for each GW event.
We point out hyperparameters in the usual population synthesis simulation context are not actually parameters related to the population,
but parameters about the evolutionary function.
This means the binary evolution functions can be constrained on an individual event basis.
We "back propagate" the posterior in the observable space to the evolutionary parameter space,
thus allowing us to study hyperparameters and theirs correlations with progenitor parameters in a data-driven manner.
Our method makes less assumptions than the traditional forward modelling approach,
which often fix the hyperparameters across the entire population.
Since we are not limited to the fixed hyperparameters assumptions, we can explore the behavior of the hyperparameters across the population much more efficiently.
While our work lays down a data analysis pathway to understand the population of GW events,
no physics can be learned without a comprehensive physical model.
We hope this letter will motivate the construction of next-generation population synthesis codes that have the following properties:
first, they should retain explicit control over the random variables so marginalizing over random variables can be done precisely and
second, they should be as automatically differentiable as possible so that exploring the evolutionary parameter space is efficient.
By combining the methods presented in this work and future, differentiable population synthesis codes,
we can explore the full parameter space of binary evolution models with future GW data.

\section{Acknowledgement}
The authors are grateful for discussions with the CCA Gravitational Wave Astronomy Group.
The Flatiron Institute is supported by the Simons Foundation.  We thank M. Zevin for helpful comments and suggestions on this manuscript.
This research has made use of data or software obtained from the Gravitational Wave Open Science Center (gw-openscience.org),
a service of LIGO Laboratory, the LIGO Scientific Collaboration, the Virgo Collaboration, and KAGRA.
LIGO Laboratory and Advanced LIGO are funded by the United States National Science Foundation (NSF) as well as
the Science and Technology Facilities Council (STFC) of the United Kingdom, the Max-Planck-Society (MPS),
and the State of Niedersachsen/Germany for support of the construction of Advanced LIGO and construction and operation of the GEO600 detector.
Additional support for Advanced LIGO was provided by the Australian Research Council. Virgo is funded, through the European Gravitational Observatory (EGO),
by the French Centre National de Recherche Scientifique (CNRS), the Italian Istituto Nazionale di Fisica Nucleare (INFN) and the Dutch Nikhef,
with contributions by institutions from Belgium, Germany, Greece, Hungary, Ireland, Japan, Monaco, Poland, Portugal, Spain.
The construction and operation of KAGRA are funded by Ministry of Education, Culture, Sports, Science and Technology (MEXT),
and Japan Society for the Promotion of Science (JSPS), National Research Foundation (NRF) and Ministry of Science and ICT (MSIT) in Korea,
Academia Sinica (AS) and the Ministry of Science and Technology (MoST) in Taiwan.

\software{
    \texttt{corner}\ \citep{corner};
    \texttt{COSMIC}\ \citep{Breivik2020};
    \texttt{julia}\ \citep{Julia-2017}
    \texttt{matplotlib}\ \citep{matplotlib};
    \texttt{numpy}\ \citep{numpy};
    \texttt{pandas}\ \citep{McKinneyPandas, reback2020pandas};
    \texttt{scipy}\ \citep{scipy}
    \texttt{seaborn}\ \citep{Waskom2021}
    \texttt{showyourwork}\ \citep{Luger2021}
}

\bibliography{bib}

\begin{thebibliography}{}
\expandafter\ifx\csname natexlab\endcsname\relax\def\natexlab#1{#1}\fi
\providecommand{\url}[1]{\href{#1}{#1}}
\providecommand{\dodoi}[1]{doi:~\href{http://doi.org/#1}{\nolinkurl{#1}}}
\providecommand{\doeprint}[1]{\href{http://ascl.net/#1}{\nolinkurl{http://ascl.net/#1}}}
\providecommand{\doarXiv}[1]{\href{https://arxiv.org/abs/#1}{\nolinkurl{https://arxiv.org/abs/#1}}}

\bibitem[{Abbott {et~al.}(2016)Abbott, Abbott, Abbott, Abernathy, Acernese,
  Ackley, Adams, Adams, Addesso, Adhikari, Adya, Affeldt, Agathos, Agatsuma,
  Aggarwal, Aguiar, Aiello, Ain, Ajith, Allen, Allocca, Altin, Anderson,
  Anderson, Arai, Arain, Araya, Arceneaux, Areeda, Arnaud, Arun, Ascenzi,
  Ashton, Ast, Aston, Astone, Aufmuth, Aulbert, Babak, Bacon, Bader, Baker,
  Baldaccini, Ballardin, Ballmer, Barayoga, Barclay, Barish, Barker, Barone,
  Barr, Barsotti, Barsuglia, Barta, Bartlett, Barton, Bartos, Bassiri, Basti,
  Batch, Baune, Bavigadda, Bazzan, Behnke, Bejger, Belczynski, Bell, Bell,
  Berger, Bergman, Bergmann, Berry, Bersanetti, Bertolini, Betzwieser, Bhagwat,
  Bhandare, Bilenko, Billingsley, Birch, Birney, Birnholtz, Biscans, Bisht,
  Bitossi, Biwer, Bizouard, Blackburn, Blair, Blair, Blair, Bloemen, Bock,
  Bodiya, Boer, Bogaert, Bogan, Bohe, Bojtos, Bond, Bondu, Bonnand, Boom, Bork,
  Boschi, Bose, Bouffanais, Bozzi, Bradaschia, Brady, Braginsky, Branchesi,
  Brau, Briant, Brillet, Brinkmann, Brisson, Brockill, Brooks, Brown, Brown,
  Brown, Buchanan, Buikema, Bulik, Bulten, Buonanno, Buskulic, Buy, Byer,
  Cabero, Cadonati, Cagnoli, Cahillane, Bustillo, Callister, Calloni, Camp,
  Cannon, Cao, Capano, Capocasa, Carbognani, Caride, Diaz, Casentini, Caudill,
  Cavagli\`a, Cavalier, Cavalieri, Cella, Cepeda, Baiardi, Cerretani, Cesarini,
  Chakraborty, Chalermsongsak, Chamberlin, Chan, Chao, Charlton,
  Chassande-Mottin, Chen, Chen, Cheng, Chincarini, Chiummo, Cho, Cho, Chow,
  Christensen, Chu, Chua, Chung, Ciani, Clara, Clark, Cleva, Coccia, Cohadon,
  Colla, Collette, Cominsky, Constancio, Conte, Conti, Cook, Corbitt, Cornish,
  Corsi, Cortese, Costa, Coughlin, Coughlin, Coulon, Countryman, Couvares,
  Cowan, Coward, Cowart, Coyne, Coyne, Craig, Creighton, Creighton, Cripe,
  Crowder, Cruise, Cumming, Cunningham, Cuoco, Canton, Danilishin, D'Antonio,
  Danzmann, Darman, Da~Silva~Costa, Dattilo, Dave, Daveloza, Davier, Davies,
  Daw, Day, De, DeBra, Debreczeni, Degallaix, De~Laurentis, Del\'eglise,
  Del~Pozzo, Denker, Dent, Dereli, Dergachev, DeRosa, De~Rosa, DeSalvo,
  Dhurandhar, D\'{\i}az, Di~Fiore, Di~Giovanni, Di~Lieto, Di~Pace, Di~Palma,
  Di~Virgilio, Dojcinoski, Dolique, Donovan, Dooley, Doravari, Douglas, Downes,
  Drago, Drever, Driggers, Du, Ducrot, Dwyer, Edo, Edwards, Effler, Eggenstein,
  Ehrens, Eichholz, Eikenberry, Engels, Essick, Etzel, Evans, Evans, Everett,
  Factourovich, Fafone, Fair, Fairhurst, Fan, Fang, Farinon, Farr, Farr,
  Favata, Fays, Fehrmann, Fejer, Feldbaum, Ferrante, Ferreira, Ferrini,
  Fidecaro, Finn, Fiori, Fiorucci, Fisher, Flaminio, Fletcher, Fong, Fournier,
  Franco, Frasca, Frasconi, Frede, Frei, Freise, Frey, Frey, Fricke, Fritschel,
  Frolov, Fulda, Fyffe, Gabbard, Gair, Gammaitoni, Gaonkar, Garufi, Gatto,
  Gaur, Gehrels, Gemme, Gendre, Genin, Gennai, George, Gergely, Germain, Ghosh,
  Ghosh, Ghosh, Giaime, Giardina, Giazotto, Gill, Glaefke, Gleason, Goetz,
  Goetz, Gondan, Gonz\'alez, Castro, Gopakumar, Gordon, Gorodetsky, Gossan,
  Gosselin, Gouaty, Graef, Graff, Granata, Grant, Gras, Gray, Greco, Green,
  Greenhalgh, Groot, Grote, Grunewald, Guidi, Guo, Gupta, Gupta, Gushwa,
  Gustafson, Gustafson, Hacker, Hall, Hall, Hammond, Haney, Hanke, Hanks,
  Hanna, Hannam, Hanson, Hardwick, Harms, Harry, Harry, Hart, Hartman, Haster,
  Haughian, Healy, Heefner, Heidmann, Heintze, Heinzel, Heitmann, Hello,
  Hemming, Hendry, Heng, Hennig, Heptonstall, Heurs, Hild, Hoak, Hodge, Hofman,
  Hollitt, Holt, Holz, Hopkins, Hosken, Hough, Houston, Howell, Hu, Huang,
  Huerta, Huet, Hughey, Husa, Huttner, Huynh-Dinh, Idrisy, Indik, Ingram, Inta,
  Isa, Isac, Isi, Islas, Isogai, Iyer, Izumi, Jacobson, Jacqmin, Jang, Jani,
  Jaranowski, Jawahar, Jim\'enez-Forteza, Johnson, Johnson-McDaniel, Jones,
  Jones, Jonker, Ju, Haris, Kalaghatgi, Kalogera, Kandhasamy, Kang, Kanner,
  Karki, Kasprzack, Katsavounidis, Katzman, Kaufer, Kaur, Kawabe, Kawazoe,
  K\'ef\'elian, Kehl, Keitel, Kelley, Kells, Kennedy, Keppel, Key,
  Khalaidovski, Khalili, Khan, Khan, Khan, Khazanov, Kijbunchoo, Kim, Kim, Kim,
  Kim, Kim, Kim, King, King, Kinzel, Kissel, Kleybolte, Klimenko, Koehlenbeck,
  Kokeyama, Koley, Kondrashov, Kontos, Koranda, Korobko, Korth, Kowalska,
  Kozak, Kringel, Krishnan, Kr\'olak, Krueger, Kuehn, Kumar, Kumar, Kuo,
  Kutynia, Kwee, Lackey, Landry, Lange, Lantz, Lasky, Lazzarini, Lazzaro,
  Leaci, Leavey, Lebigot, Lee, Lee, Lee, Lee, Lenon, Leonardi, Leong, Leroy,
  Letendre, Levin, Levine, Li, Libson, Littenberg, Lockerbie, Logue, Lombardi,
  London, Lord, Lorenzini, Loriette, Lormand, Losurdo, Lough, Lousto, Lovelace,
  L\"uck, Lundgren, Luo, Lynch, Ma, MacDonald, Machenschalk, MacInnis, Macleod,
  Maga\~na Sandoval, Magee, Mageswaran, Majorana, Maksimovic, Malvezzi, Man,
  Mandel, Mandic, Mangano, Mansell, Manske, Mantovani, Marchesoni, Marion,
  M\'arka, M\'arka, Markosyan, Maros, Martelli, Martellini, Martin, Martin,
  Martynov, Marx, Mason, Masserot, Massinger, Masso-Reid, Matichard, Matone,
  Mavalvala, Mazumder, Mazzolo, McCarthy, McClelland, McCormick, McGuire,
  McIntyre, McIver, McManus, McWilliams, Meacher, Meadors, Meidam, Melatos,
  Mendell, Mendoza-Gandara, Mercer, Merilh, Merzougui, Meshkov, Messenger,
  Messick, Meyers, Mezzani, Miao, Michel, Middleton, Mikhailov, Milano, Miller,
  Millhouse, Minenkov, Ming, Mirshekari, Mishra, Mitra, Mitrofanov,
  Mitselmakher, Mittleman, Moggi, Mohan, Mohapatra, Montani, Moore, Moore,
  Moraru, Moreno, Morriss, Mossavi, Mours, Mow-Lowry, Mueller, Mueller, Muir,
  Mukherjee, Mukherjee, Mukherjee, Mukund, Mullavey, Munch, Murphy, Murray,
  Mytidis, Nardecchia, Naticchioni, Nayak, Necula, Nedkova, Nelemans, Neri,
  Neunzert, Newton, Nguyen, Nielsen, Nissanke, Nitz, Nocera, Nolting,
  Normandin, Nuttall, Oberling, Ochsner, O'Dell, Oelker, Ogin, Oh, Oh, Ohme,
  Oliver, Oppermann, Oram, O'Reilly, O'Shaughnessy, Ott, Ottaway, Ottens,
  Overmier, Owen, Pai, Pai, Palamos, Palashov, Palomba, Pal-Singh, Pan, Pan,
  Pankow, Pannarale, Pant, Paoletti, Paoli, Papa, Paris, Parker, Pascucci,
  Pasqualetti, Passaquieti, Passuello, Patricelli, Patrick, Pearlstone,
  Pedraza, Pedurand, Pekowsky, Pele, Penn, Perreca, Pfeiffer, Phelps, Piccinni,
  Pichot, Pickenpack, Piergiovanni, Pierro, Pillant, Pinard, Pinto, Pitkin,
  Poeld, Poggiani, Popolizio, Post, Powell, Prasad, Predoi, Premachandra,
  Prestegard, Price, Prijatelj, Principe, Privitera, Prix, Prodi, Prokhorov,
  Puncken, Punturo, Puppo, P\"urrer, Qi, Qin, Quetschke, Quintero,
  Quitzow-James, Raab, Rabeling, Radkins, Raffai, Raja, Rakhmanov, Ramet,
  Rapagnani, Raymond, Razzano, Re, Read, Reed, Regimbau, Rei, Reid, Reitze,
  Rew, Reyes, Ricci, Riles, Robertson, Robie, Robinet, Rocchi, Rolland,
  Rollins, Roma, Romano, Romano, Romanov, Romie, Rosi\ifmmode~\acute{n}\else
  \'{n}\fi{}ska, Rowan, R\"udiger, Ruggi, Ryan, Sachdev, Sadecki, Sadeghian,
  Salconi, Saleem, Salemi, Samajdar, Sammut, Sampson, Sanchez, Sandberg,
  Sandeen, Sanders, Sanders, Sassolas, Sathyaprakash, Saulson, Sauter, Savage,
  Sawadsky, Schale, Schilling, Schmidt, Schmidt, Schnabel, Schofield,
  Sch\"onbeck, Schreiber, Schuette, Schutz, Scott, Scott, Sellers, Sengupta,
  Sentenac, Sequino, Sergeev, Serna, Setyawati, Sevigny, Shaddock, Shaffer,
  Shah, Shahriar, Shaltev, Shao, Shapiro, Shawhan, Sheperd, Shoemaker,
  Shoemaker, Siellez, Siemens, Sigg, Silva, Simakov, Singer, Singer, Singh,
  Singh, Singhal, Sintes, Slagmolen, Smith, Smith, Smith, Smith, Son, Sorazu,
  Sorrentino, Souradeep, Srivastava, Staley, Steinke, Steinlechner,
  Steinlechner, Steinmeyer, Stephens, Stevenson, Stone, Strain, Straniero,
  Stratta, Strauss, Strigin, Sturani, Stuver, Summerscales, Sun, Sutton,
  Swinkels, Szczepa\ifmmode~\acute{n}\else \'{n}\fi{}czyk, Tacca, Talukder,
  Tanner, T\'apai, Tarabrin, Taracchini, Taylor, Theeg, Thirugnanasambandam,
  Thomas, Thomas, Thomas, Thorne, Thorne, Thrane, Tiwari, Tiwari, Tokmakov,
  Tomlinson, Tonelli, Torres, Torrie, T\"oyr\"a, Travasso, Traylor, Trifir\`o,
  Tringali, Trozzo, Tse, Turconi, Tuyenbayev, Ugolini, Unnikrishnan, Urban,
  Usman, Vahlbruch, Vajente, Valdes, Vallisneri, van Bakel, van Beuzekom,
  van~den Brand, Van Den~Broeck, Vander-Hyde, van~der Schaaf, van Heijningen,
  van Veggel, Vardaro, Vass, Vas\'uth, Vaulin, Vecchio, Vedovato, Veitch,
  Veitch, Venkateswara, Verkindt, Vetrano, Vicer\'e, Vinciguerra, Vine, Vinet,
  Vitale, Vo, Vocca, Vorvick, Voss, Vousden, Vyatchanin, Wade, Wade, Wade,
  Waldman, Walker, Wallace, Walsh, Wang, Wang, Wang, Wang, Wang, Ward, Ward,
  Warner, Was, Weaver, Wei, Weinert, Weinstein, Weiss, Welborn, Wen,
  We\ss{}els, Westphal, Wette, Whelan, Whitcomb, White, Whiting, Wiesner,
  Wilkinson, Willems, Williams, Williams, Williamson, Willis, Willke, Wimmer,
  Winkelmann, Winkler, Wipf, Wiseman, Wittel, Woan, Worden, Wright, Wu, Yablon,
  Yakushin, Yam, Yamamoto, Yancey, Yap, Yu, Yvert, Zadro\ifmmode~\dot{z}\else
  \.{z}\fi{}ny, Zangrando, Zanolin, Zendri, Zevin, Zhang, Zhang, Zhang, Zhang,
  Zhao, Zhou, Zhou, Zhu, Zucker, Zuraw, \& Zweizig}]{GW150914}
Abbott, B.~P., Abbott, R., Abbott, T.~D., {et~al.} 2016, Phys. Rev. Lett., 116,
  061102, \dodoi{10.1103/PhysRevLett.116.061102}

\bibitem[{{Abbott} {et~al.}(2018){Abbott}, {Abbott}, {Abbott}, {Abernathy},
  {Acernese}, {Ackley}, {Adams}, {Adams}, {et~al.}}]{Abbott2018}
{Abbott}, B.~P., {Abbott}, R., {Abbott}, T.~D., {et~al.} 2018, Living Reviews
  in Relativity, 21, 3, \dodoi{10.1007/s41114-018-0012-9}

\bibitem[{Abbott {et~al.}(2021)}]{LIGOScientific:2019lzm}
Abbott, R., {et~al.} 2021, SoftwareX, 13, 100658,
  \dodoi{10.1016/j.softx.2021.100658}

\bibitem[{{Acernese} {et~al.}(2015){Acernese}, {Agathos}, {Agatsuma}, {Aisa},
  {Allemandou}, {Allocca}, {Amarni}, {Astone}, {Balestri}, {Ballardin}, \&
  et~al.}]{Acernese2015}
{Acernese}, F., {Agathos}, M., {Agatsuma}, K., {et~al.} 2015, Classical and
  Quantum Gravity, 32, 024001, \dodoi{10.1088/0264-9381/32/2/024001}

\bibitem[{{Acernese} {et~al.}(2019){Acernese}, {Agathos}, {Aiello}, {Allocca},
  {Amato}, {Ansoldi}, {Antier}, {Ar{\`e}ne}, {Arnaud}, {Ascenzi}, \&
  et~al.}]{Acernese2019}
{Acernese}, F., {Agathos}, M., {Aiello}, L., {et~al.} 2019, \prl, 123, 231108,
  \dodoi{10.1103/PhysRevLett.123.231108}

\bibitem[{{Akutsu} {et~al.}(2021){Akutsu}, {Ando}, {Arai}, {Arai}, {Araki},
  {Araya}, {Aritomi}, {Aso}, {Bae}, {et~al.}}]{Akutsu2021}
{Akutsu}, T., {Ando}, M., {Arai}, K., {et~al.} 2021, Progress of Theoretical
  and Experimental Physics, 2021, 05A101, \dodoi{10.1093/ptep/ptaa125}

\bibitem[{{Ali-Ha{\"\i}moud} {et~al.}(2017){Ali-Ha{\"\i}moud}, {Kovetz}, \&
  {Kamionkowski}}]{AliHaimoud2017}
{Ali-Ha{\"\i}moud}, Y., {Kovetz}, E.~D., \& {Kamionkowski}, M. 2017, \prd, 96,
  123523, \dodoi{10.1103/PhysRevD.96.123523}

\bibitem[{{Andrews} {et~al.}(2021){Andrews}, {Cronin}, {Kalogera}, {Berry}, \&
  {Zezas}}]{Andrews2021}
{Andrews}, J.~J., {Cronin}, J., {Kalogera}, V., {Berry}, C. P.~L., \& {Zezas},
  A. 2021, \apjl, 914, L32, \dodoi{10.3847/2041-8213/ac00a6}

\bibitem[{{Andrews} {et~al.}(2018){Andrews}, {Zezas}, \&
  {Fragos}}]{Andrews2018}
{Andrews}, J.~J., {Zezas}, A., \& {Fragos}, T. 2018, \apjs, 237, 1,
  \dodoi{10.3847/1538-4365/aaca30}

\bibitem[{{Antonini} \& {Rasio}(2016)}]{Antonini2016}
{Antonini}, F., \& {Rasio}, F.~A. 2016, \apj, 831, 187,
  \dodoi{10.3847/0004-637X/831/2/187}

\bibitem[{{Antonini} {et~al.}(2017){Antonini}, {Toonen}, \&
  {Hamers}}]{Antonini2017}
{Antonini}, F., {Toonen}, S., \& {Hamers}, A.~S. 2017, \apj, 841, 77,
  \dodoi{10.3847/1538-4357/aa6f5e}

\bibitem[{{Ashton} {et~al.}(2019){Ashton}, {H{\"u}bner}, {Lasky}, {Talbot},
  {Ackley}, {Biscoveanu}, {Chu}, {Divakarla}, {Easter}, {Goncharov}, {Hernandez
  Vivanco}, {Harms}, {Lower}, {Meadors}, {Melchor}, {Payne}, {Pitkin},
  {Powell}, {Sarin}, {Smith}, \& {Thrane}}]{Ashton2019}
{Ashton}, G., {H{\"u}bner}, M., {Lasky}, P.~D., {et~al.} 2019, \apjs, 241, 27,
  \dodoi{10.3847/1538-4365/ab06fc}

\bibitem[{{Askar} {et~al.}(2017){Askar}, {Szkudlarek}, {Gondek-Rosi{\'n}ska},
  {Giersz}, \& {Bulik}}]{Askar2017}
{Askar}, A., {Szkudlarek}, M., {Gondek-Rosi{\'n}ska}, D., {Giersz}, M., \&
  {Bulik}, T. 2017, \mnras, 464, L36, \dodoi{10.1093/mnrasl/slw177}

\bibitem[{{Aso} {et~al.}(2013){Aso}, {Michimura}, {Somiya}, {Ando}, {Miyakawa},
  {Sekiguchi}, {Tatsumi}, \& {Yamamoto}}]{Aso2013}
{Aso}, Y., {Michimura}, Y., {Somiya}, K., {et~al.} 2013, \prd, 88, 043007,
  \dodoi{10.1103/PhysRevD.88.043007}

\bibitem[{{Banerjee}(2017)}]{Banerjee2017}
{Banerjee}, S. 2017, \mnras, 467, 524, \dodoi{10.1093/mnras/stw3392}

\bibitem[{{Bavera} {et~al.}(2020){Bavera}, {Fragos}, {Qin}, {Zapartas},
  {Neijssel}, {Mandel}, {Batta}, {Gaebel}, {Kimball}, \&
  {Stevenson}}]{Bavera2020}
{Bavera}, S.~S., {Fragos}, T., {Qin}, Y., {et~al.} 2020, \aap, 635, A97,
  \dodoi{10.1051/0004-6361/201936204}

\bibitem[{{Bavera} {et~al.}(2021){Bavera}, {Fragos}, {Zevin}, {Berry},
  {Marchant}, {Andrews}, {Coughlin}, {Dotter}, {Kovlakas}, {Misra},
  {Serra-Perez}, {Qin}, {Rocha}, {Rom{\'a}n-Garza}, {Tran}, \&
  {Zapartas}}]{Bavera2021}
{Bavera}, S.~S., {Fragos}, T., {Zevin}, M., {et~al.} 2021, \aap, 647, A153,
  \dodoi{10.1051/0004-6361/202039804}

\bibitem[{{Belczynski} {et~al.}(2004){Belczynski}, {Bulik}, \&
  {Rudak}}]{Belczynski2004}
{Belczynski}, K., {Bulik}, T., \& {Rudak}, B. 2004, \apjl, 608, L45,
  \dodoi{10.1086/422172}

\bibitem[{{Belczynski} {et~al.}(2002){Belczynski}, {Kalogera}, \&
  {Bulik}}]{Belczynski2002}
{Belczynski}, K., {Kalogera}, V., \& {Bulik}, T. 2002, \apj, 572, 407,
  \dodoi{10.1086/340304}

\bibitem[{{Belczynski} {et~al.}(2008){Belczynski}, {Kalogera}, {Rasio}, {Taam},
  {Zezas}, {Bulik}, {Maccarone}, \& {Ivanova}}]{Belczynski2008}
{Belczynski}, K., {Kalogera}, V., {Rasio}, F.~A., {et~al.} 2008, \apjs, 174,
  223, \dodoi{10.1086/521026}

\bibitem[{{Belczynski} {et~al.}(2016){Belczynski}, {Repetto}, {Holz},
  {O'Shaughnessy}, {Bulik}, {Berti}, {Fryer}, \& {Dominik}}]{Belczynski2016}
{Belczynski}, K., {Repetto}, S., {Holz}, D.~E., {et~al.} 2016, \apj, 819, 108,
  \dodoi{10.3847/0004-637X/819/2/108}

\bibitem[{Bezanson {et~al.}(2017)Bezanson, Edelman, Karpinski, \&
  Shah}]{Julia-2017}
Bezanson, J., Edelman, A., Karpinski, S., \& Shah, V.~B. 2017, SIAM {R}eview,
  59, 65, \dodoi{10.1137/141000671}

\bibitem[{{Bird} {et~al.}(2016){Bird}, {Cholis}, {Mu{\~n}oz},
  {Ali-Ha{\"\i}moud}, {Kamionkowski}, {Kovetz}, {Raccanelli}, \&
  {Riess}}]{Bird2016}
{Bird}, S., {Cholis}, I., {Mu{\~n}oz}, J.~B., {et~al.} 2016, \prl, 116, 201301,
  \dodoi{10.1103/PhysRevLett.116.201301}

\bibitem[{{Breivik} {et~al.}(2020){Breivik}, {Coughlin}, {Zevin}, {Rodriguez},
  {Kremer}, {Ye}, {Andrews}, {Kurkowski}, {Digman}, {Larson}, \&
  {Rasio}}]{Breivik2020}
{Breivik}, K., {Coughlin}, S., {Zevin}, M., {et~al.} 2020, \apj, 898, 71,
  \dodoi{10.3847/1538-4357/ab9d85}

\bibitem[{{Broekgaarden} {et~al.}(2021){Broekgaarden}, {Berger}, {Stevenson},
  {Justham}, {Mandel}, {Chru{\'s}li{\'n}ska}, {van Son}, {Wagg},
  {Vigna-G{\'o}mez}, {de Mink}, {Chattopadhyay}, \&
  {Neijssel}}]{Broekgaarden2021}
{Broekgaarden}, F.~S., {Berger}, E., {Stevenson}, S., {et~al.} 2021, arXiv
  e-prints, arXiv:2112.05763.
\newblock \doarXiv{2112.05763}

\bibitem[{{Buikema} {et~al.}(2020){Buikema}, {Cahillane}, {Mansell}, {Blair},
  {Abbott}, {Adams}, {Adhikari}, {Ananyeva}, {Appert}, {et~al.}}]{Buikema2020}
{Buikema}, A., {Cahillane}, C., {Mansell}, G.~L., {et~al.} 2020, \prd, 102,
  062003, \dodoi{10.1103/PhysRevD.102.062003}

\bibitem[{{Chattopadhyay} {et~al.}(2022){Chattopadhyay}, {Hurley}, {Stevenson},
  \& {Raidani}}]{Chattopadhyay2022}
{Chattopadhyay}, D., {Hurley}, J., {Stevenson}, S., \& {Raidani}, A. 2022,
  \mnras, \dodoi{10.1093/mnras/stac1163}

\bibitem[{{de Mink} \& {Mandel}(2016)}]{deMink2016}
{de Mink}, S.~E., \& {Mandel}, I. 2016, \mnras, 460, 3545,
  \dodoi{10.1093/mnras/stw1219}

\bibitem[{{Di Carlo} {et~al.}(2020){Di Carlo}, {Mapelli}, {Giacobbo}, {Spera},
  {Bouffanais}, {Rastello}, {Santoliquido}, {Pasquato}, {Ballone}, {Trani},
  {Torniamenti}, \& {Haardt}}]{DiCarlo2020}
{Di Carlo}, U.~N., {Mapelli}, M., {Giacobbo}, N., {et~al.} 2020, \mnras, 498,
  495, \dodoi{10.1093/mnras/staa2286}

\bibitem[{{Dominik} {et~al.}(2012){Dominik}, {Belczynski}, {Fryer}, {Holz},
  {Berti}, {Bulik}, {Mandel}, \& {O'Shaughnessy}}]{Dominik2012}
{Dominik}, M., {Belczynski}, K., {Fryer}, C., {et~al.} 2012, \apj, 759, 52,
  \dodoi{10.1088/0004-637X/759/1/52}

\bibitem[{{Downing} {et~al.}(2010){Downing}, {Benacquista}, {Giersz}, \&
  {Spurzem}}]{Downing2010}
{Downing}, J.~M.~B., {Benacquista}, M.~J., {Giersz}, M., \& {Spurzem}, R. 2010,
  \mnras, 407, 1946, \dodoi{10.1111/j.1365-2966.2010.17040.x}

\bibitem[{{Farmer} {et~al.}(2019){Farmer}, {Renzo}, {de Mink}, {Marchant}, \&
  {Justham}}]{Farmer2019}
{Farmer}, R., {Renzo}, M., {de Mink}, S.~E., {Marchant}, P., \& {Justham}, S.
  2019, \apj, 887, 53, \dodoi{10.3847/1538-4357/ab518b}

\bibitem[{{Ford} \& {McKernan}(2021)}]{Ford2021}
{Ford}, K.~E.~S., \& {McKernan}, B. 2021, arXiv e-prints, arXiv:2109.03212.
\newblock \doarXiv{2109.03212}

\bibitem[{Foreman-Mackey(2016)}]{corner}
Foreman-Mackey, D. 2016, The Journal of Open Source Software, 1, 24,
  \dodoi{10.21105/joss.00024}

\bibitem[{{Fragione} \& {Kocsis}(2019)}]{Fragione2019}
{Fragione}, G., \& {Kocsis}, B. 2019, \mnras, 486, 4781,
  \dodoi{10.1093/mnras/stz1175}

\bibitem[{{Fryer} {et~al.}(2012){Fryer}, {Belczynski}, {Wiktorowicz},
  {Dominik}, {Kalogera}, \& {Holz}}]{Fryer2012}
{Fryer}, C.~L., {Belczynski}, K., {Wiktorowicz}, G., {et~al.} 2012, \apj, 749,
  91, \dodoi{10.1088/0004-637X/749/1/91}

\bibitem[{{Fuller} \& {Ma}(2019)}]{Fuller2019}
{Fuller}, J., \& {Ma}, L. 2019, \apjl, 881, L1,
  \dodoi{10.3847/2041-8213/ab339b}

\bibitem[{{Gallegos-Garcia} {et~al.}(2021){Gallegos-Garcia}, {Berry},
  {Marchant}, \& {Kalogera}}]{Gallego-Garcia2021}
{Gallegos-Garcia}, M., {Berry}, C. P.~L., {Marchant}, P., \& {Kalogera}, V.
  2021, \apj, 922, 110, \dodoi{10.3847/1538-4357/ac2610}

\bibitem[{{Hunter}(2007)}]{matplotlib}
{Hunter}, J.~D. 2007, Computing in Science and Engineering, 9, 90,
  \dodoi{10.1109/MCSE.2007.55}

\bibitem[{{Inayoshi} {et~al.}(2017){Inayoshi}, {Hirai}, {Kinugawa}, \&
  {Hotokezaka}}]{Inayoshi2017}
{Inayoshi}, K., {Hirai}, R., {Kinugawa}, T., \& {Hotokezaka}, K. 2017, \mnras,
  468, 5020, \dodoi{10.1093/mnras/stx757}

\bibitem[{{Inayoshi} {et~al.}(2016){Inayoshi}, {Kashiyama}, {Visbal}, \&
  {Haiman}}]{Inayoshi2016}
{Inayoshi}, K., {Kashiyama}, K., {Visbal}, E., \& {Haiman}, Z. 2016, \mnras,
  461, 2722, \dodoi{10.1093/mnras/stw1431}

\bibitem[{{Ivanova} \& {Taam}(2004)}]{Ivanova2004}
{Ivanova}, N., \& {Taam}, R.~E. 2004, \apj, 601, 1058, \dodoi{10.1086/380561}

\bibitem[{Jones {et~al.}(2001)Jones, Oliphant, Peterson, {et~al.}}]{scipy}
Jones, E., Oliphant, T., Peterson, P., {et~al.} 2001, {SciPy}: Open source
  scientific tools for {Python}.
\newblock \url{http://www.scipy.org/}

\bibitem[{{Kinugawa} {et~al.}(2014){Kinugawa}, {Inayoshi}, {Hotokezaka},
  {Nakauchi}, \& {Nakamura}}]{Kinugawa2014}
{Kinugawa}, T., {Inayoshi}, K., {Hotokezaka}, K., {Nakauchi}, D., \&
  {Nakamura}, T. 2014, \mnras, 442, 2963, \dodoi{10.1093/mnras/stu1022}

\bibitem[{{LIGO Scientific Collaboration} {et~al.}(2015){LIGO Scientific
  Collaboration}, {Aasi}, {Abbott}, {Abbott}, {Abbott}, {Abernathy}, {Ackley},
  {Adams}, {Adams}, {Addesso}, \& et~al.}]{LIGOScientificCollaboration2015}
{LIGO Scientific Collaboration}, {Aasi}, J., {Abbott}, B.~P., {et~al.} 2015,
  Classical and Quantum Gravity, 32, 074001,
  \dodoi{10.1088/0264-9381/32/7/074001}

\bibitem[{{Luger} {et~al.}(2021){Luger}, {Bedell}, {Foreman-Mackey},
  {Crossfield}, {Zhao}, \& {Hogg}}]{Luger2021}
{Luger}, R., {Bedell}, M., {Foreman-Mackey}, D., {et~al.} 2021, arXiv e-prints,
  arXiv:2110.06271.
\newblock \doarXiv{2110.06271}

\bibitem[{{Mandel} \& {Broekgaarden}(2022)}]{Mandel2022}
{Mandel}, I., \& {Broekgaarden}, F.~S. 2022, Living Reviews in Relativity, 25,
  1, \dodoi{10.1007/s41114-021-00034-3}

\bibitem[{{Mandel} \& {de Mink}(2016)}]{Mandel2016}
{Mandel}, I., \& {de Mink}, S.~E. 2016, \mnras, 458, 2634,
  \dodoi{10.1093/mnras/stw379}

\bibitem[{{Mapelli} {et~al.}(2022){Mapelli}, {Bouffanais}, {Santoliquido},
  {Arca Sedda}, \& {Artale}}]{Mapelli2022}
{Mapelli}, M., {Bouffanais}, Y., {Santoliquido}, F., {Arca Sedda}, M., \&
  {Artale}, M.~C. 2022, \mnras, 511, 5797, \dodoi{10.1093/mnras/stac422}

\bibitem[{{Marchant} {et~al.}(2016){Marchant}, {Langer}, {Podsiadlowski},
  {Tauris}, \& {Moriya}}]{Marchant2016}
{Marchant}, P., {Langer}, N., {Podsiadlowski}, P., {Tauris}, T.~M., \&
  {Moriya}, T.~J. 2016, \aap, 588, A50, \dodoi{10.1051/0004-6361/201628133}

\bibitem[{{McKernan} {et~al.}(2020){McKernan}, {Ford}, \&
  {O'Shaughnessy}}]{McKernan2020}
{McKernan}, B., {Ford}, K.~E.~S., \& {O'Shaughnessy}, R. 2020, \mnras, 498,
  4088, \dodoi{10.1093/mnras/staa2681}

\bibitem[{{McKernan} {et~al.}(2018){McKernan}, {Ford}, {Bellovary}, {Leigh},
  {Haiman}, {Kocsis}, {Lyra}, {Mac Low}, {Metzger}, {O'Dowd}, {Endlich}, \&
  {Rosen}}]{McKernan2018}
{McKernan}, B., {Ford}, K.~E.~S., {Bellovary}, J., {et~al.} 2018, \apj, 866,
  66, \dodoi{10.3847/1538-4357/aadae5}

\bibitem[{{Miller} \& {Lauburg}(2009)}]{Miller2009}
{Miller}, M.~C., \& {Lauburg}, V.~M. 2009, \apj, 692, 917,
  \dodoi{10.1088/0004-637X/692/1/917}

\bibitem[{{Ng} {et~al.}(2021){Ng}, {Vitale}, {Farr}, \& {Rodriguez}}]{Ng2021}
{Ng}, K. K.~Y., {Vitale}, S., {Farr}, W.~M., \& {Rodriguez}, C.~L. 2021, \apjl,
  913, L5, \dodoi{10.3847/2041-8213/abf8be}

\bibitem[{{O'Leary} {et~al.}(2006){O'Leary}, {Rasio}, {Fregeau}, {Ivanova}, \&
  {O'Shaughnessy}}]{OLeary2006}
{O'Leary}, R.~M., {Rasio}, F.~A., {Fregeau}, J.~M., {Ivanova}, N., \&
  {O'Shaughnessy}, R. 2006, \apj, 637, 937, \dodoi{10.1086/498446}

\bibitem[{pandas~development team(2020)}]{reback2020pandas}
pandas~development team, T. 2020, pandas-dev/pandas: Pandas, 1.1.1,  Zenodo,
  \dodoi{10.5281/zenodo.3509134}

\bibitem[{{Planck Collaboration} {et~al.}(2020){Planck Collaboration},
  {Aghanim}, {Akrami}, {Ashdown}, {Aumont}, {Baccigalupi}, {Ballardini},
  {Banday}, {Barreiro}, {Bartolo}, {Basak}, {Battye}, {Benabed}, {Bernard},
  {Bersanelli}, {Bielewicz}, {Bock}, {Bond}, {Borrill}, {Bouchet}, {Boulanger},
  {Bucher}, {Burigana}, {Butler}, {Calabrese}, {Cardoso}, {Carron},
  {Challinor}, {Chiang}, {Chluba}, {Colombo}, {Combet}, {Contreras}, {Crill},
  {Cuttaia}, {de Bernardis}, {de Zotti}, {Delabrouille}, {Delouis}, {Di
  Valentino}, {Diego}, {Dor{\'e}}, {Douspis}, {Ducout}, {Dupac}, {Dusini},
  {Efstathiou}, {Elsner}, {En{\ss}lin}, {Eriksen}, {Fantaye}, {Farhang},
  {Fergusson}, {Fernandez-Cobos}, {Finelli}, {Forastieri}, {Frailis},
  {Fraisse}, {Franceschi}, {Frolov}, {Galeotta}, {Galli}, {Ganga},
  {G{\'e}nova-Santos}, {Gerbino}, {Ghosh}, {Gonz{\'a}lez-Nuevo}, {G{\'o}rski},
  {Gratton}, {Gruppuso}, {Gudmundsson}, {Hamann}, {Handley}, {Hansen},
  {Herranz}, {Hildebrandt}, {Hivon}, {Huang}, {Jaffe}, {Jones}, {Karakci},
  {Keih{\"a}nen}, {Keskitalo}, {Kiiveri}, {Kim}, {Kisner}, {Knox},
  {Krachmalnicoff}, {Kunz}, {Kurki-Suonio}, {Lagache}, {Lamarre}, {Lasenby},
  {Lattanzi}, {Lawrence}, {Le Jeune}, {Lemos}, {Lesgourgues}, {Levrier},
  {Lewis}, {Liguori}, {Lilje}, {Lilley}, {Lindholm}, {L{\'o}pez-Caniego},
  {Lubin}, {Ma}, {Mac{\'\i}as-P{\'e}rez}, {Maggio}, {Maino}, {Mandolesi},
  {Mangilli}, {Marcos-Caballero}, {Maris}, {Martin}, {Martinelli},
  {Mart{\'\i}nez-Gonz{\'a}lez}, {Matarrese}, {Mauri}, {McEwen}, {Meinhold},
  {Melchiorri}, {Mennella}, {Migliaccio}, {Millea}, {Mitra},
  {Miville-Desch{\^e}nes}, {Molinari}, {Montier}, {Morgante}, {Moss}, {Natoli},
  {N{\o}rgaard-Nielsen}, {Pagano}, {Paoletti}, {Partridge}, {Patanchon},
  {Peiris}, {Perrotta}, {Pettorino}, {Piacentini}, {Polastri}, {Polenta},
  {Puget}, {Rachen}, {Reinecke}, {Remazeilles}, {Renzi}, {Rocha}, {Rosset},
  {Roudier}, {Rubi{\~n}o-Mart{\'\i}n}, {Ruiz-Granados}, {Salvati}, {Sandri},
  {Savelainen}, {Scott}, {Shellard}, {Sirignano}, {Sirri}, {Spencer},
  {Sunyaev}, {Suur-Uski}, {Tauber}, {Tavagnacco}, {Tenti}, {Toffolatti},
  {Tomasi}, {Trombetti}, {Valenziano}, {Valiviita}, {Van Tent}, {Vibert},
  {Vielva}, {Villa}, {Vittorio}, {Wandelt}, {Wehus}, {White}, {White},
  {Zacchei}, \& {Zonca}}]{Planck2018}
{Planck Collaboration}, {Aghanim}, N., {Akrami}, Y., {et~al.} 2020, \aap, 641,
  A6, \dodoi{10.1051/0004-6361/201833910}

\bibitem[{{Portegies Zwart} \& {McMillan}(2000)}]{PortegiesZwart2000}
{Portegies Zwart}, S.~F., \& {McMillan}, S. L.~W. 2000, \apjl, 528, L17,
  \dodoi{10.1086/312422}

\bibitem[{{Riley} {et~al.}(2022){Riley}, {Agrawal}, {Barrett}, {Boyett},
  {Broekgaarden}, {Chattopadhyay}, {Gaebel}, {Gittins}, {Hirai}, {Howitt},
  {Justham}, {Khandelwal}, {Kummer}, {Lau}, {Mandel}, {de Mink}, {Neijssel},
  {Riley}, {van Son}, {Stevenson}, {Vigna-G{\'o}mez}, {Vinciguerra}, {Wagg},
  {Willcox}, \& {Team Compas}}]{Riley2022}
{Riley}, J., {Agrawal}, P., {Barrett}, J.~W., {et~al.} 2022, \apjs, 258, 34,
  \dodoi{10.3847/1538-4365/ac416c}

\bibitem[{{Rodriguez} {et~al.}(2016){Rodriguez}, {Chatterjee}, \&
  {Rasio}}]{Rodriguez2016}
{Rodriguez}, C.~L., {Chatterjee}, S., \& {Rasio}, F.~A. 2016, \prd, 93, 084029,
  \dodoi{10.1103/PhysRevD.93.084029}

\bibitem[{{Rodriguez} {et~al.}(2015){Rodriguez}, {Morscher}, {Pattabiraman},
  {Chatterjee}, {Haster}, \& {Rasio}}]{Rodriguez2015}
{Rodriguez}, C.~L., {Morscher}, M., {Pattabiraman}, B., {et~al.} 2015, \prl,
  115, 051101, \dodoi{10.1103/PhysRevLett.115.051101}

\bibitem[{{Rodriguez} {et~al.}(2019){Rodriguez}, {Zevin}, {Amaro-Seoane},
  {Chatterjee}, {Kremer}, {Rasio}, \& {Ye}}]{Rodriguez2019}
{Rodriguez}, C.~L., {Zevin}, M., {Amaro-Seoane}, P., {et~al.} 2019, \prd, 100,
  043027, \dodoi{10.1103/PhysRevD.100.043027}

\bibitem[{{Romero-Shaw} {et~al.}(2020){Romero-Shaw}, {Talbot}, {Biscoveanu},
  {D'Emilio}, {Ashton}, {Berry}, {Coughlin}, {Galaudage}, {Hoy}, {H{\"u}bner},
  {Phukon}, {Pitkin}, {Rizzo}, {Sarin}, {Smith}, {Stevenson}, {Vajpeyi},
  {Ar{\`e}ne}, {Athar}, {Banagiri}, {Bose}, {Carney}, {Chatziioannou}, {Clark},
  {Colleoni}, {Cotesta}, {Edelman}, {Estell{\'e}s}, {Garc{\'\i}a-Quir{\'o}s},
  {Ghosh}, {Green}, {Haster}, {Husa}, {Keitel}, {Kim}, {Hernandez-Vivanco},
  {Maga{\~n}a Hernandez}, {Karathanasis}, {Lasky}, {De Lillo}, {Lower},
  {Macleod}, {Mateu-Lucena}, {Miller}, {Millhouse}, {Morisaki}, {Oh},
  {Ossokine}, {Payne}, {Powell}, {Pratten}, {P{\"u}rrer}, {Ramos-Buades},
  {Raymond}, {Thrane}, {Veitch}, {Williams}, {Williams}, \&
  {Xiao}}]{Romero-Shaw2020}
{Romero-Shaw}, I.~M., {Talbot}, C., {Biscoveanu}, S., {et~al.} 2020, \mnras,
  499, 3295, \dodoi{10.1093/mnras/staa2850}

\bibitem[{{Samsing} {et~al.}(2014){Samsing}, {MacLeod}, \&
  {Ramirez-Ruiz}}]{Samsing2014}
{Samsing}, J., {MacLeod}, M., \& {Ramirez-Ruiz}, E. 2014, \apj, 784, 71,
  \dodoi{10.1088/0004-637X/784/1/71}

\bibitem[{{Secunda} {et~al.}(2020){Secunda}, {Bellovary}, {Mac Low}, {Ford},
  {McKernan}, {Leigh}, {Lyra}, {S{\'a}ndor}, \& {Adorno}}]{Secunda2020}
{Secunda}, A., {Bellovary}, J., {Mac Low}, M.-M., {et~al.} 2020, \apj, 903,
  133, \dodoi{10.3847/1538-4357/abbc1d}

\bibitem[{{Silsbee} \& {Tremaine}(2017)}]{Silsbee2017}
{Silsbee}, K., \& {Tremaine}, S. 2017, \apj, 836, 39,
  \dodoi{10.3847/1538-4357/aa5729}

\bibitem[{{Stevenson} {et~al.}(2017){Stevenson}, {Vigna-G{\'o}mez}, {Mandel},
  {Barrett}, {Neijssel}, {Perkins}, \& {de Mink}}]{Stevenson2017}
{Stevenson}, S., {Vigna-G{\'o}mez}, A., {Mandel}, I., {et~al.} 2017, Nature
  Communications, 8, 14906, \dodoi{10.1038/ncomms14906}

\bibitem[{{Tanikawa} {et~al.}(2021){Tanikawa}, {Susa}, {Yoshida}, {Trani}, \&
  {Kinugawa}}]{Tanikawa2021}
{Tanikawa}, A., {Susa}, H., {Yoshida}, T., {Trani}, A.~A., \& {Kinugawa}, T.
  2021, \apj, 910, 30, \dodoi{10.3847/1538-4357/abe40d}

\bibitem[{{Tanikawa} {et~al.}(2022){Tanikawa}, {Yoshida}, {Kinugawa}, {Trani},
  {Hosokawa}, {Susa}, \& {Omukai}}]{Tanikawa2022}
{Tanikawa}, A., {Yoshida}, T., {Kinugawa}, T., {et~al.} 2022, \apj, 926, 83,
  \dodoi{10.3847/1538-4357/ac4247}

\bibitem[{{The LIGO Scientific Collaboration} {et~al.}(2021){The LIGO
  Scientific Collaboration}, {the Virgo Collaboration}, {the KAGRA
  Collaboration}, {Abbott}, {Abbott}, {Acernese}, {Ackley}, {Adams},
  {Adhikari}, {Adhikari}, \& et~al.}]{GWTC-3}
{The LIGO Scientific Collaboration}, {the Virgo Collaboration}, {the KAGRA
  Collaboration}, {et~al.} 2021, arXiv e-prints, arXiv:2111.03606.
\newblock \doarXiv{2111.03606}

\bibitem[{{Tse} {et~al.}(2019){Tse}, {Yu}, {Kijbunchoo}, {Fernandez-Galiana},
  {Dupej}, {Barsotti}, {Blair}, {Brown}, {Dwyer}, {Effler}, \&
  et~al.}]{Tse2019}
{Tse}, M., {Yu}, H., {Kijbunchoo}, N., {et~al.} 2019, \prl, 123, 231107,
  \dodoi{10.1103/PhysRevLett.123.231107}

\bibitem[{{van der Walt} {et~al.}(2011){van der Walt}, {Colbert}, \&
  {Varoquaux}}]{numpy}
{van der Walt}, S., {Colbert}, S.~C., \& {Varoquaux}, G. 2011, Computing in
  Science and Engineering, 13, 22, \dodoi{10.1109/MCSE.2011.37}

\bibitem[{{van Son} {et~al.}(2022){van Son}, {de Mink}, {Callister}, {Justham},
  {Renzo}, {Wagg}, {Broekgaarden}, {Kummer}, {Pakmor}, \&
  {Mandel}}]{VanSon2022}
{van Son}, L.~A.~C., {de Mink}, S.~E., {Callister}, T., {et~al.} 2022, \apj,
  931, 17, \dodoi{10.3847/1538-4357/ac64a3}

\bibitem[{{Veitch} {et~al.}(2015){Veitch}, {Raymond}, {Farr}, {Farr}, {Graff},
  {Vitale}, {Aylott}, {Blackburn}, {Christensen}, {Coughlin}, {Del Pozzo},
  {Feroz}, {Gair}, {Haster}, {Kalogera}, {Littenberg}, {Mandel},
  {O'Shaughnessy}, {Pitkin}, {Rodriguez}, {R{\"o}ver}, {Sidery}, {Smith}, {Van
  Der Sluys}, {Vecchio}, {Vousden}, \& {Wade}}]{Veitch2015}
{Veitch}, J., {Raymond}, V., {Farr}, B., {et~al.} 2015, \prd, 91, 042003,
  \dodoi{10.1103/PhysRevD.91.042003}

\bibitem[{{Vigna-G{\'o}mez} {et~al.}(2021){Vigna-G{\'o}mez}, {Toonen},
  {Ramirez-Ruiz}, {Leigh}, {Riley}, \& {Haster}}]{VignaGomez2021}
{Vigna-G{\'o}mez}, A., {Toonen}, S., {Ramirez-Ruiz}, E., {et~al.} 2021, \apjl,
  907, L19, \dodoi{10.3847/2041-8213/abd5b7}

\bibitem[{{Vitale} {et~al.}(2019){Vitale}, {Farr}, {Ng}, \&
  {Rodriguez}}]{Vitale2019}
{Vitale}, S., {Farr}, W.~M., {Ng}, K. K.~Y., \& {Rodriguez}, C.~L. 2019, \apjl,
  886, L1, \dodoi{10.3847/2041-8213/ab50c0}

\bibitem[{Waskom(2021)}]{Waskom2021}
Waskom, M.~L. 2021, Journal of Open Source Software, 6, 3021,
  \dodoi{10.21105/joss.03021}

\bibitem[{{W}es {M}c{K}inney(2010)}]{McKinneyPandas}
{W}es {M}c{K}inney. 2010, in {P}roceedings of the 9th {P}ython in {S}cience
  {C}onference, ed. {S}t\'efan van~der {W}alt \& {J}arrod {M}illman, 56 -- 61,
  \dodoi{10.25080/Majora-92bf1922-00a}

\bibitem[{{Wong} {et~al.}(2021){Wong}, {Breivik}, {Kremer}, \&
  {Callister}}]{Wong2021}
{Wong}, K. W.~K., {Breivik}, K., {Kremer}, K., \& {Callister}, T. 2021, \prd,
  103, 083021, \dodoi{10.1103/PhysRevD.103.083021}

\bibitem[{{Woosley}(2017)}]{Woosley2017}
{Woosley}, S.~E. 2017, \apj, 836, 244, \dodoi{10.3847/1538-4357/836/2/244}

\bibitem[{{Zevin} {et~al.}(2020){Zevin}, {Spera}, {Berry}, \&
  {Kalogera}}]{Zevin2020}
{Zevin}, M., {Spera}, M., {Berry}, C. P.~L., \& {Kalogera}, V. 2020, \apjl,
  899, L1, \dodoi{10.3847/2041-8213/aba74e}

\bibitem[{{Zevin} {et~al.}(2021){Zevin}, {Bavera}, {Berry}, {Kalogera},
  {Fragos}, {Marchant}, {Rodriguez}, {Antonini}, {Holz}, \&
  {Pankow}}]{Zevin2021}
{Zevin}, M., {Bavera}, S.~S., {Berry}, C. P.~L., {et~al.} 2021, \apj, 910, 152,
  \dodoi{10.3847/1538-4357/abe40e}

\bibitem[{{Ziosi} {et~al.}(2014){Ziosi}, {Mapelli}, {Branchesi}, \&
  {Tormen}}]{Ziosi2014}
{Ziosi}, B.~M., {Mapelli}, M., {Branchesi}, M., \& {Tormen}, G. 2014, \mnras,
  441, 3703, \dodoi{10.1093/mnras/stu824}

\end{thebibliography}

\end{document}